\documentstyle[aps,epsfig]{revtex}
\begin{document}
\title{Microscopic chaos and diffusion}
\author{C. P. Dettmann and E. G. D. Cohen}
\address{Rockefeller University, 1230 York Ave, New York NY 10021, USA}
\date{\today}
\maketitle

\begin{abstract}
We investigate the connections between microscopic chaos,
defined on a dynamical level and arising from collisions between molecules,
and diffusion, characterized by a mean square displacement proportional
to the time.  We use a number of models involving a single particle moving
in two dimensions and colliding with fixed scatterers.  We find that a number
of microscopically nonchaotic models exhibit diffusion,
and that the standard methods of
chaotic time series analysis are ill suited to the problem of distinguishing
between chaotic and nonchaotic microscopic dynamics.
However, we show that periodic orbits
play an important role in our models, in that their different properties
in our chaotic and nonchaotic models can be used to distinguish them
at the level of time series analysis, and in systems with absorbing boundaries.
Our findings are relevant to experiments aimed at verifying the existence of
chaoticity and related dynamical properties on a microscopic level
in diffusive systems.
\end{abstract}

\section{Introduction}
It is generally assumed that diffusion in macroscopic systems is a result
of chaos on a microscopic scale, following Einstein's 1905 explanation
for Brownian motion, that is, the movement of a colloidal particle due
to thermal fluctuations in the surrounding fluid.  
We describe chaos, diffusion and related properties in detail in
Sec.~\ref{s:fund}.  We define
microscopic chaos (usually shortened to just ``chaos'') in terms of
unpredictability as quantified by positive Lyapunov exponents or a positive
Kolmogorov-Sinai entropy;%
\footnote{Specifically, a positive Lyapunov exponent of a dynamical system
means that initially similar states of the system as measured by distance
in phase space separate exponentially fast; the closely related
property of positive Kolmogorov-Sinai entropy means that a finite amount
of information per unit time is needed to construct the future phase space
trajectory of the system, knowing the infinite past trajectory to an
arbitrary (but finite) resolution.}
later we will find that properties based on periodic orbits also provide a
useful description of chaotic aspects of the microscopic dynamics.
Note that we restrict the use of
the term ``chaotic'' to its usual meaning, that is a positive Kolmogorov-Sinai
entropy; while the properties of periodic orbits are important, we do not
classify systems as chaotic or nonchaotic based on these properties.
We call a system diffusive
when the mean square displacement of a particle is proportional
to the time, or a distribution of particles satisfies the diffusion
equation. 

A recent experiment of Gaspard et al~\cite{GBFSGDC}, described below,
purports to show that the diffusion of a Brownian particle is due to
microscopic chaos.  While we believe that Brownian motion (including
both the Brownian particle and the solvent) is most likely chaotic,
our simulations using nonchaotic models, in a brief Comment~\cite{DCvB}
and in greater detail here lead to the same results as found in the
experiment, so that no experimental proof of microscopic chaos has
been given in~\cite{GBFSGDC}.
Here we explore the question of what kind of experimental measurements or
data analysis might be required to identify microscopic chaotic dynamics,
and the connection between microscopic chaos and diffusion.

\begin{figure}
\begin{picture}(280,280)(-50,0)
\put(40,10){\epsfig{file=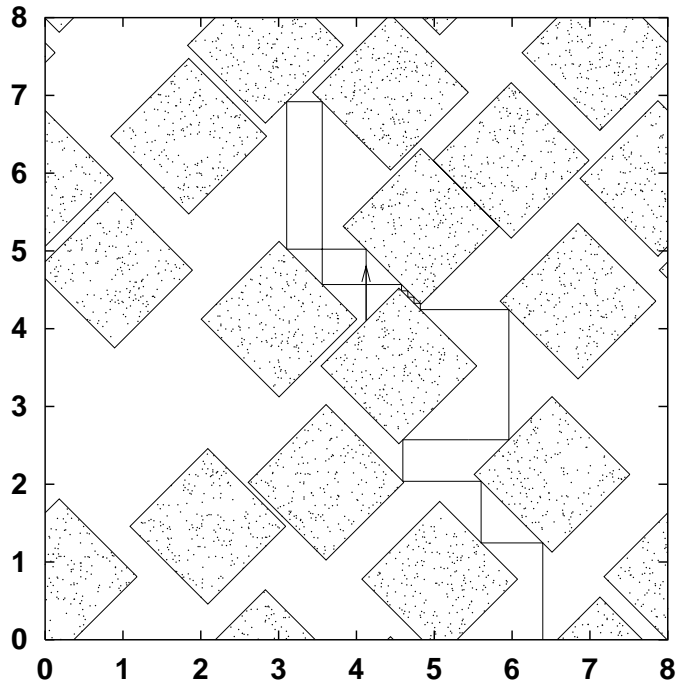, width=280pt}}
\end{picture}
\caption{The fixed orientation wind-tree model.  There are periodic boundary
conditions, so in the notation of section~\protect\ref{s:not} this is FP8.
\label{f:fix}}
\end{figure}

\begin{figure}
\begin{picture}(280,280)(-50,0)
\put(40,10){\epsfig{file=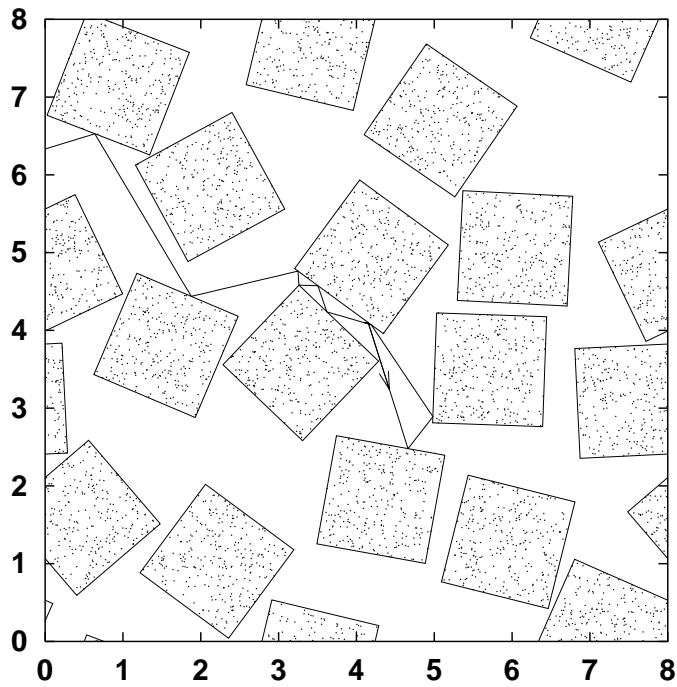, width=280pt}}
\end{picture}
\caption{The randomly oriented wind-tree model, with notation RP8.
\label{f:ran}}
\end{figure}

\begin{figure}
\begin{picture}(280,280)(-50,0)
\put(40,10){\epsfig{file=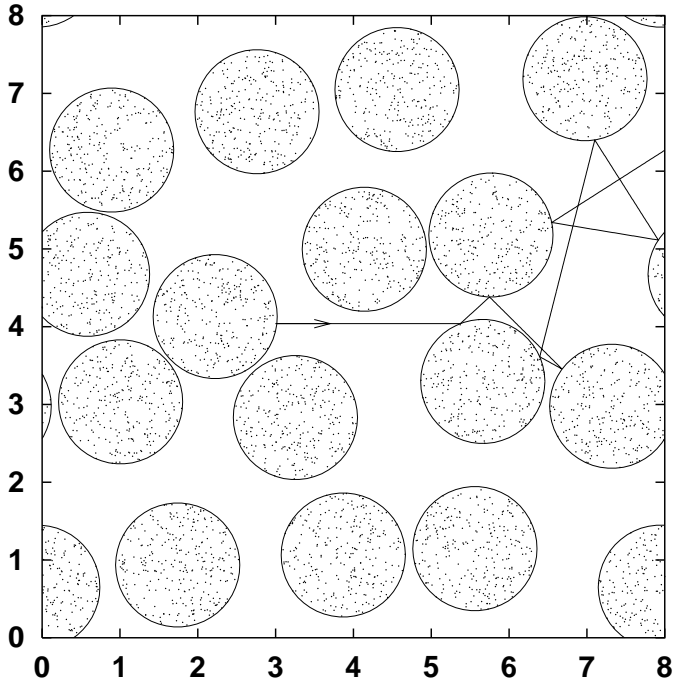, width=280pt}}
\end{picture}
\caption{The Lorentz gas, with notation LP8.
\label{f:lor}}
\end{figure}

We consider generalizations of models of diffusion due to Lorentz and
Ehrenfest (Figs.~\ref{f:fix}-\ref{f:lor}), where a single point particle
undergoes elastic collisions with a fixed arrangement of circular or square
scatterers in two dimensions, respectively.
Collisions with the circular scatterers of the (chaotic) Lorentz gas lead to
exponential separation of nearby trajectories, that is, a positive
Lyapunov exponent, while collisions with the square scatterers of the
(nonchaotic) Ehrenfest model lead to at most linear separation of
nearby trajectories, and the Lyapunov exponents are all zero. 
Actually, we consider models with three types of scatterers, all of which are
non-overlapping: circles as in the Lorentz gas, squares oriented such
that their diagonals are parallel to the coordinate axes and with
only four particle velocity directions as in the Ehrenfest wind-tree model,
and squares of arbitrary orientations with arbitrary particle velocity
directions as a generalization of the Ehrenfest model.

For each of these three models, we consider the following three cases:
an infinite number of randomly placed fixed scatterers as
in the original models of Lorentz and Ehrenfest, a small number of fixed
scatterers in an elementary cell subject to periodic boundary conditions,
and an arrangement of a finite number of fixed scatterers enclosed by
absorbing boundaries.  Note that a model with periodic boundary conditions
can be viewed as an infinite periodic system.  For the purposes of studying
properties such as the mean free time, it is better to view the
model as a finite system with periodic boundaries, but to compute the
diffusion coefficient in terms of the growth of the mean square displacement
with time, it is necessary to view the model as an infinite periodic system.

A link between chaos and diffusion involves a fundamental question of
statistical mechanics, since one has to make a connection between the
microscopic and macroscopic behavior of large systems.  The models we
consider here are very special from a physical point of view, but very
attractive from a mathematical point of view, because there is only
one moving particle, rather than a large number as in, for example, Brownian
motion.  These can be considered bona fide statistical mechanical models
if the large number of scatterers are included, as long as their lack of motion
is irrelevant to the questions we ask.  We believe that our models
incorporate the essential features needed for a discussion of microscopic
chaos and diffusion.

We now describe the experiment and analysis of Gaspard et al. in more detail.
The position of a Brownian particle in a fluid was measured at regular
intervals of $1/60\;\;{\rm s}$.  The experimental time series data was
then interpreted using standard techniques of chaotic time series analysis,
suggesting a positive lower bound on the Kolmogorov-Sinai entropy,
hence microscopic chaos.  The method they used, due to Cohen and
Procaccia~\cite{CP} was adapted from an approach pioneered by Grassberger and
Procaccia~\cite{GP}.  In this method one analyzes the distribution of
recurrences,  {\em i.e.} instances
where the system approximately retraces
part of its previous trajectory in phase space for a certain length of time,
leading to a determination of the Kolmogorov-Sinai entropy.
There is detailed mathematical justification for the method (under
assumptions such as the length of the data set being sufficient to
approximate infinite time limits, see Sec.~\ref{s:GP}), but the idea
is very simple: recurrences give useful information
about the predictability of the system.  If long recurrences occur very
often, it is easy to predict the future of the system from previous
instances similar to the recent part of the trajectory, so the system
has a small or zero Kolmogorov-Sinai entropy,
whereas a rapid decay of the frequency with length of the recurrences
indicates a high degree of microscopic chaos.  Thus, the Cohen-Procaccia
method used by Gaspard can be used to calculate the Kolmogorov-Sinai
entropy from a time series in principle, but certain mathematical conditions
and limits apply, sometimes restricting its applicability in practice.
Gaspard et al. concluded from their analysis that the Kolmogorov-Sinai
entropy of the system containing the Brownian particle was positive,
and hence that the microscopic dynamics was chaotic.

Subsequently, we showed~\cite{DCvB} that the same approach applied to an 
``identical'' time series generated by a numerical simulation of the nonchaotic
(infinite, fixed orientation) Ehrenfest model yielded virtually
identical results.  This is surprising, since the collisions with the flat
sides of square scatterers do not lead to positive Lyapunov exponents,
as described above.  In Ref.~\cite{DCvB} we attributed the discrepancy to the
physical issue of time scales --- the time interval between measurements
($1/60\;\;{\rm s}$) was vastly greater than the typical collision times of
the Brownian particle with the solvent particles in the fluid (approximately
$10^{-12}\;\;{\rm s}$).  While this is certainly an experimental problem,
it leaves open the
question of whether in principle a similar analysis with a much higher
resolution could ever prove from experimental data that the microscopic
dynamics is chaotic or not.

One aim of the current work is to shed light on this, and related questions.
We perform the same Cohen-Procaccia analysis on all the models mentioned above,
finding that the same results are obtained even for a model with a periodic 
array of squares.  This rules out the possibility that the apparently positive
value of the Kolmogorov-Sinai entropy is due to a loss of information
associated with the randomness of the many scatterers.  We discuss the
relevant time scales in our models, finding a single microscopic time scale,
and we can then qualitatively explain the
results of the Cohen-Procaccia algorithm when the time step is shorter or
longer than
this time.  We conclude that due to the rarity of recurrences in a diffusive
system, the determination of chaoticity in such a system requires vastly
longer time series than are practical for experimental or computational
work, even when the measurements are taken at microscopic time and distance
scales.

At that point we return to the question of how microscopic dynamics manifests
itself in a time series, as well as in the macroscopic behavior.
We develop a new method  which we call ``almost periodic recurrences'' that
selects a particular class of recurrences different from those used in the
Cohen-Procaccia method.  Our approach is designed to distinguish systems
based on their periodic orbit properties, which are related to, but not
equivalent to the usual definition of chaos as positive Kolmogorov-Sinai
entropy.  We find that, in contrast to the Cohen-Procaccia method,
this method can
distinguish between the chaotic Lorentz model and the nonchaotic
Ehrenfest model and thus reveals at least one way in which microscopic
chaos is manifest, although on macroscopic scales both models exhibit
diffusion.  The basic idea of our method is simple: almost periodic
recurrences are related to periodic orbits which have
very different properties in chaotic and nonchaotic systems, so a search
for periodic orbits using recurrences may distinguish between the two classes
of systems.
The periodic orbits of chaotic systems (such as the Lorentz gas) are
exponentially unstable, so many repetitions close to a periodic orbit are
unlikely.  In contrast,
the periodic orbits of nonchaotic systems (for example our wind-tree models)
can be power law unstable, allowing many repetitions.

The markedly different properties of periodic orbits in the Lorentz and
Ehrenfest models leads to the following important observations about diffusion
in finite geometries.
Solutions of the diffusion equation with absorbing boundaries exhibit
exponential decay, corresponding to the probability of a particle
remaining in the system for a given time.  The randomly oriented wind-tree
model has periodic orbits with power law escape, so that the escape from
the whole system cannot be exponential at long times.  The fixed orientation
wind-tree model has no periodic orbits at all, so the particle cannot remain in
the system longer than a fixed maximum time.  Both of these are examples
of situations in which the infinite model satisfies the diffusion equation
by having a Gaussian distribution with the mean square displacement
proportional to the time, but the corresponding finite model does not
satisfy the diffusion equation. 

The outline of this paper is as follows:  first we introduce our models
(Sec.~\ref{s:mod}), and discuss in detail their relevant time scales and
chaotic and diffusive properties (Sec.~\ref{s:fund}).
Then we discuss the Cohen-Procaccia method and its inability
to distinguish chaotic from nonchaotic dynamics in our case (Sec.~\ref{s:GP}).
Finally we introduce our alternative ``almost periodic recurrence''
method and discuss the properties of
the periodic orbits (Sec.~\ref{s:ISR}), and their consequences for finite
systems (Sec.~\ref{s:esc}).  We conclude with a discussion of our results
and some open questions (Sec.\ref{s:conc}).

\section{The models}\label{s:mod}
\subsection{Definitions}
\label{s:desc}
The Ehrenfest wind-tree model describes a point ``wind'' particle
moving in straight lines in the plane punctuated by elastic collisions
with fixed square scatterers, the ``trees''.  In the original model
(Fig.~\ref{f:fix}), the
trees are oriented with their diagonals along the $x$ and $y$ axes,
and the wind particle moves with a fixed velocity in only four possible
directions, along these axes.
The trees are located at random positions, with a given number density, and
not overlapping.  There is an overlapping version we will not consider; it
exhibits anomalous (sub)-diffusion and non-Gaussian
distributions~\cite{HC}.

Secondly, we consider the case of randomly oriented scatterers
(Fig.~\ref{f:ran}),
which allows all possible wind particle directions.  In both the fixed
and randomly oriented cases all
Lyapunov exponents are zero, because a collision from a flat scatterer
causes only a linear separation of nearby trajectories; it does not cause
the exponential separation associated with convex curved scatterers.

Thirdly, we consider the (two dimensional) Lorentz gas (Fig.~\ref{f:lor}),
that is, a model
where the scatterers are circular, and we choose the area and number density
of the scatterers to correspond to the wind-tree models.
This model is known to have normal diffusive
properties~\cite{vB,Gaspard} and is used here as a comparison to examine
the effects of positive versus zero Lyapunov exponents. 

\subsection{Numerical details}\label{s:comp}
For the numerical simulations, we take the velocity of the wind particle to
be $1$.  For the wind-tree models we take the side length $l$ equal to
$\sqrt{2}$.  This is done for computational convenience; space is divided
into unit squares (of side length 1) aligned with the coordinate axes, so that
at most one scatterer can be contained in each unit square. 
For the Lorentz gas the circular scatterers have a radius $R=\sqrt{2/\pi}$
to give them the same area $l^2=2$ as the squares.
In all cases the total area considered has a number of unit squares, $L$, up to
$3500$ in both $x$ and $y$ directions, and periodic boundary conditions are
used.  To simulate an infinite system, a large value of $L$ (up to 3500)
is used; for
the maximum time of $10^6$ time units the particle never travels far enough
to sample the periodic boundary conditions.  To simulate a periodic system,
a small size such as $L=4$ is used; we always consider the case of finite
horizon, so that the time between collisions is bounded.  To simulate a
system with absorbing boundaries, an intermediate size such as $L=20$ is
used; the absorbing boundaries are one distance unit inside the edge of the
system to avoid the possibility of the particle colliding with a scatterer
as well as with its periodic image. 

We use a number of scatterers $N$ equal to $L^2/4$.  This means that the
number density $n=N/L^2$ is $1/4$ in all cases and the proportion of the total
area covered by the scatterers is $\rho=N(l/L)^2=1/2$.  Thus we have an
intermediate density of scatterers, which is most convenient from the
point of view of the simulations and also the time scales.  For a significantly
lower density, an infinite horizon is much more likely in all but the largest
systems.  For a significantly higher density (limited by a maximum of
$\rho=1$) the Metropolis algorithm (see below) used to position the scatterers
is very much slower.  Also, both for low and high density systems, there is an
additional time scale (see section~\ref{s:time}), which complicates
the analysis. 

We now describe the method used to position the fixed scatterers.
Each configuration of random positions and orientations is obtained
from a version of the Metropolis algorithm, used in~\cite{WL}.
The square scatterers of the wind-tree models are placed initially
at points belonging to a square lattice defined such that
they do not overlap, that is, with their centers at integer coordinates
$(m,n)$ such that $m+n$ is even, and their orientation as in Fig.~\ref{f:fix}.
Not every lattice site is then occupied, depending on the number density.
\footnote{We use a random initial placement of the scatterers on the lattice
to accelerate the convergence of the shifting algorithm, described next.
The random placement leads to large scale fluctuations in density.  These
fluctuations would occur anyway as a result of the small shifts, but only
after a very large number of iterations.}
Each scatterer is shifted and rotated a small random amount in turn, typically
a few tenths of a distance unit and of order ten degrees respectively.
If the shift causes an overlap, the move is rejected and the previous
configuration is used.  The procedure is repeated with different shifts
and rotations.  A large fixed number of shifts are attempted,
sufficient for reasonable measures of the correlation between scatterers
to have long converged to their final values. 

If the circles of the Lorentz gas are placed on the same
lattice as the squares they overlap slightly; at the chosen density
of $\rho=1/2$ there is enough freedom to allow them to find non-overlapping
locations, so that the final configuration is non-overlapping, as we could
check.  At higher densities circles may not find non-overlapping locations,
and a close packed triangular lattice should then be used instead.

\subsection{Notation}\label{s:not}
We now define the notation we will use to distinguish between the
various models.
There are a very large number of possible parameters that could appear
in general, referring to the shape of the scatterers (squares, circles,
etc.), different densities and different shaped boundary conditions
to name a few, but for conciseness we limit the notation to include
only those models we use here, in particular keeping the areal density
always at $\rho=1/2$.
Using square brackets [,] to denote different cases, our notation is
[F,R,L][$\infty$,[P,A]$L$].  The first symbol [F,R,L] denotes the type and
orientation of the scatterers, F for fixed oriented squares, R for
randomly oriented squares and L for circles (the Lorentz gas).  The
next symbol [$\infty$,P,A] gives information about the boundary conditions,
either $\infty$, that is, no boundaries, or P for periodic boundary
conditions
or A for absorbing boundaries.  The symbols P and A are followed by the
size of the system, $L$, which is an even integer.  Thus the original
Ehrenfest model is denoted F$\infty$, while a periodic Lorentz gas with $L=8$
(containing $N=L^2/4=16$ scatterers at the density used in our simulations)
is denoted LP8 (see Fig.~\ref{f:lor}).  The latter model is generated
by random shifts of the
scatterers as above, ensuring that periodic images of the scatterers
do not overlap.  When no confusion can arise, we will sometimes
refer to classes of models with a simplified notation, for example we denote
all randomly oriented wind-tree models as R, and all Lorentz models with
absorbing boundaries as LA. 

\section{Fundamental properties}\label{s:fund}
\subsection{Time scales}\label{s:time}
This section collects a number of fundamental properties and results that
form a basis for sections~\ref{s:GP},~\ref{s:ISR} and~\ref{s:esc}.
First we consider the important issue of {\em time scales}, which puts the time
series analysis methods of sections~\ref{s:GP} and~\ref{s:ISR} into a physical
context. Then we consider {\em chaotic properties}, giving known results and
conjectures for our models; all of the following sections use this.
After this we study {\em diffusive properties}, obtaining some new results.
These are of direct importance to Sec.~\ref{s:esc}; the time series analysis
methods are applied only to our diffusive models,
\footnote{We include here one model exhibiting superdiffusion,
see Sec.~\ref{s:diff}.}
 but the diffusion
is somewhat incidental as the methods can be applied to arbitrary time
series.  At the end of this section we relate the chaotic and diffusive
properties of our models.

One of our difficulties of the Gaspard et al experiment~\cite{GBFSGDC}
is that the
interval between their measurements (1/60 s) is so much greater than the
relevant time scale ($10^{-12}\;\;{\rm s}$) of the dynamics, that is the
time interval determined
by collisions between the Brownian particle and the other particles in the
fluid.  It is thus important to clarify the issue of what time scales are
relevant in our models, so that the simulation results can be put in context.
There are three relevant time scales here, the time taken for the particle
to traverse the length of a scatterer (defined to be of order one time unit;
see Sec.~\ref{s:comp}), the mean free time between collisions
$\bar{\tau}$, and the time at which the particle begins to notice the finite
size $L$ of the system (when it is not infinite).  It turns out
(see Sec.~\ref{s:diff})
that all the random and periodic models except FP have a well defined
diffusion coefficient $D$, so in these cases the time taken to reach the
boundary is of order $L^2/D$.

The mean free time can be calculated exactly (see the Appendix)
and is given by $\bar{\tau}_F=1$, $\bar{\tau}_R=\pi/(2\sqrt{2})\approx 1.111$
and $\bar{\tau}_L=\sqrt{\pi/2}\approx 1.253$ for the F, R, and L models
respectively, independent of the size of the system (ignoring the case of
absorbing boundaries).  We have checked these values numerically, and the
results agree within their uncertainties of about $0.002$.
It is clear that all the time scales in these models are of order one time
unit, except those defined by the boundary.  This means that a time step
of one time unit should be sufficient to observe effects due to the
microscopic chaos in the time series analysis methods of
sections~\ref{s:GP} and~\ref{s:ISR}.

\subsection{Microscopic dynamical properties}\label{s:chaos}
We began with the notion of microscopic chaos as the presence of a
positive Lyapunov exponent arising from collisions with strictly convex
scatterers, or as a positive Kolmogorov-Sinai entropy quantifying the
lack of predictability of a chaotic system.  We now want to make these
ideas more precise and clarify what is known and what is conjectured about
our models.

From a mathematical point of view, there are a large number
of dynamical properties that a system may
exhibit~\cite{ER}.  Those of interest to us are

\begin{enumerate}
\item Ergodicity.  We need ergodicity for the time series
analysis methods of Secs.~\ref{s:GP} and~\ref{s:ISR}; that is, a single
long trajectory is supposed to sample the dynamics of the whole phase
space.  We either know or conjecture that ergodicity holds in all of
our models (see below).%
\item Decay of velocity correlations.  We note that
Eqs.~(\ref{e:GK},\ref{e:GKB}) below give the diffusion and Burnett
coefficients as integrals over velocity autocorrelation functions:
the diffusion coefficient involves two-time correlations, while the
Burnett coefficient involves four-time correlations.
For these coefficients to be defined the appropriate correlations
must decay sufficiently quickly.
\footnote{Mixing (which implies ergodicity) is equivalent to the
statement that all two time correlation functions decay, not just velocity
correlations.  It is neither
necessary nor sufficient for the existence of the diffusion coefficient:
it is not necessary because correlations other than velocity correlations
need not decay in a diffusive system, and it is not sufficient because
the velocity correlation could decay too slowly for the integral to converge.}
Numerical evidence for the existence
or nonexistence of these coefficients for our models is given in
Sec.~\ref{s:diff}.
\item A positive Lyapunov exponent, corresponding to exponential separation
of initially close trajectories, follows from the shape of the scatterers.
The Lorentz gas has a positive Lyapunov exponent due to its convex
scatterers, while the wind-tree models have all zero Lyapunov exponents
due to their (piecewise) flat scatterers.
\footnote{We are not interested in the exact value of the Lyapunov exponent
in the Lorentz gas (although it is easy to calculate numerically) because
there is no general relation linking it to, for example, the diffusion
coefficient.  For example, in our wind-tree models the Lyapunov exponent
is zero, but the diffusion coefficient remains positive.  The same is
true for the KS entropy.}
\item A positive Kolmogorov-Sinai entropy, which is our definition of
chaos, and which the Cohen-Procaccia method
(see Sec.~\ref{s:GP}) is designed to compute.  We either know or
conjecture that the KS entropy is equal to the sum of the positive
Lyapunov exponents in our models (Pesin's theorem, see below).
Actually there is at most one positive Lyapunov exponent in these systems.
\item Periodic orbit properties sensitive to chaoticity are described 
in Sec.~\ref{s:ISR} and used in Secs.~\ref{s:ISR} and~\ref{s:esc}.
\end{enumerate}

We now briefly discuss what is known about our models regarding these
dynamical properties.
The periodic Lorentz gas has exponential decay of (all two time)
correlations~\cite{Young}, which implies ergodicity and the convergence of
the integral in~(\ref{e:GK}) below.
For~(\ref{e:GKB}) see Ref.~\cite{CD}.  The periodic wind-tree models are
generically
\footnote{The term generic here denotes a positive measure of scatterer
configurations, as opposed to initial positions of the particle.}
ergodic~\cite{Gutkin}.  Pesin's theorem holds for all of these [F,R,L]P
models, see Refs.~\cite{ER,W} for the Lorentz gas and Ref.~\cite{Gutkin}
for the wind-tree models.

\begin{figure}
\begin{picture}(300,320)(-70,0)
\put(0,20){\epsfig{file=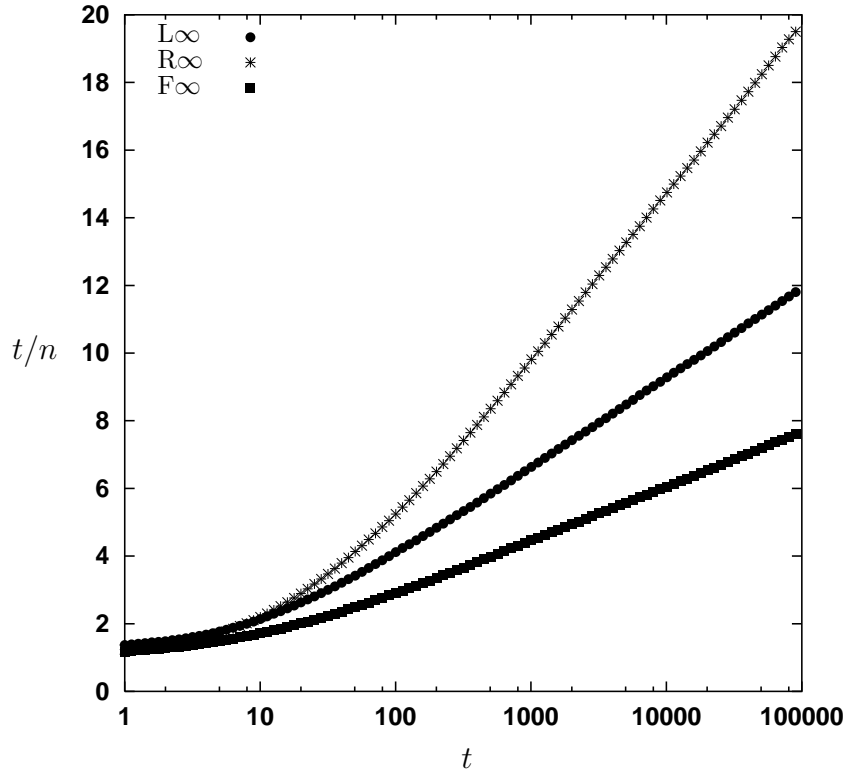, width=300pt}}
\put(155,10){\large $t$}
\put(-15,165){\large $t/n$}
\put(40,285){L$\infty$}
\put(40,275){R$\infty$}
\put(40,265){F$\infty$}
\end{picture}
\caption{Numerical evidence for conjectures applying to the infinite 
models [F,R,L]$\infty$.  If $n$ is the average number of
{\em different} scatterers hit by the particle in time $t$, we expect that
$n$ should be roughly proportional to $t/\log t$ for large $t$.  Here we
plot $t/n$ as a function of $t$, which should thus be a straight line on a
linear-log scale, as is observed.
\label{f:stat}}
\end{figure}

The infinite models are much more difficult to treat rigorously because
the phase space is not compact.  In order to make reasonable conjectures
about the above properties, we use the solution of the diffusion
equation~(\ref{e:Gauss}) below, which is verified numerically in
Sec.~\ref{s:diff} for the models [F,R,L]$\infty$, to argue that the
number of different scatterers hit by the particle in time $t$ is proportional
to $t/\log t$, as follows.  In two dimensions,
the probability density of the particle being at any point decays as $1/t$.
This means that in a time $t$, the particle comes close (say, within a radius
$\epsilon$) to the point a number of times proportional to the integral of
$1/t$, that is $\log t$.  If the point corresponds to the location of a
scatterer, we could say that the particle collides with each scatterer
a number of times proportional to $\log t$.  In this case the total number of
different scatterers hit by the particle in time $t$ might be expected to grow
proportional to $t/\log t$, which is what we find numerically in
Fig.~\ref{f:stat}.

These observations have the following consequences for the chaotic
properties of the infinite models.  Because the particle returns to each
scatterer an infinite number of times, we expect that it passes arbitrarily
close to every point in phase space, that is, the system is ergodic.
The KS entropy gives the amount of information per unit time required
to predict the trajectory knowing its infinite past.  The unpredictability
has two sources, the instability associated with the collisions in the
Lorentz gas (that is, the positive Lyapunov exponent), and the random
positions of the scatterers with which the particle has not yet collided
in all the models.  The number of different scatterers hit by the particle
grows as $t/\log t$, so the rate decreases to zero as $1/\log t$.  We
conclude from this argument that the random positions of the scatterers
do not contribute to the KS entropy, although results at finite time
might suggest otherwise, given that $1/\log t$ decays so slowly.  In other
words, Pesin's theorem is satisfied for these infinite models also.
\footnote{The argument hinges on the solution of the diffusion equation
(Eq.~(\ref{e:diff}) below) in two dimensions.  In three dimensions, the
density decays as $t^{-3/2}$ which is integrable at infinity,
so that the particle collides with each scatterer only a finite number
of times and thus does not sample the whole phase space.  In addition,
there is a contribution to the KS entropy from the random positions of the
scatterers, thus violating Pesin's theorem.  There is no contradiction
here; the proofs of Pesin's theorem all demand that the phase space be compact.}

We have discussed a number of dynamical properties with regard to our
models. The chaotic Lorentz gas satisfies more of these properties than
the nonchaotic wind-tree models.  Chaos is defined here by positive KS
entropy and is clearly relevant fro diffusion, yet we show in
subsequent sections that the periodic orbit
properties also appear to be relevant to diffusion.  The conclusion
discusses an example of a model which is chaotic as it has positive KS
entropy, but has periodic orbit properties and diffusive properties more
similar to the nonchaotic wind-tree models.  Although we have a
substantial number of relevant numerical results below, a full
understanding of the dynamical properties (KS entropy, periodic orbit   
properties or perhaps others) most closely related to diffusion still
eludes us.
 
\subsection{Macroscopic diffusion}\label{s:diff}
Now we discuss diffusive properties.  We begin with the diffusion
equation, and then relate it to the mean square displacement of a particle.
The diffusion equation
\begin{equation}\label{e:diff}
\frac{\partial P}{\partial t}=D\nabla^2 P
\end{equation}
for a probability density $P({\bf x},t)$ with a constant diffusion
coefficient $D$ is linear.  Thus its general solution is an integral over
Green's functions given by the solution for an initial Dirac delta distribution
$P({\bf x},0)=\delta({\bf x}-{\bf x}_0)$, that is,
\begin{equation}\label{e:Gauss}
P({\bf x},t)=(4\pi Dt)^{-d/2}
\exp\left[-\frac{({\bf x}-{\bf x}_0)^2}{4Dt}\right]
\end{equation}
is the conditional probability density that a particle initially at the point
${\bf x}_0$ will be at the point ${\bf x}$ a time $t$ later; this is therefore
a 2-time probability distribution (or density) function.  The spatial
dimension $d=2$ in our case.

The diffusion equation is known to hold for the probability density
of a particle undergoing deterministic dynamics in a number of systems
including the Lorentz gas~\cite{Gaspard}.  In the diffusion equation
a macroscopic approximation is
used, implying that $P({\bf x},t)$ is an average over space and time 
scales large compared to the characteristic microscopic lengths and times
of the dynamics.  The mean square displacement of the particle in the
$x$-direction after a macroscopic time $t$ is obtained from (\ref{e:Gauss}),
\begin{equation}
\langle \Delta x^2 \rangle=\int \Delta x^2 P({\bf x},t)d{\bf x}
=2Dt\label{e:msd}
\end{equation}
where $\Delta x=x-x_0$.  This is the Einstein relation for diffusion.  Note
that $x=x(t)$ is now an explicit function of time.  Similarly, the fourth
and sixth cumulants are
\begin{eqnarray}
\langle \Delta x^4 \rangle_c\equiv
\langle \Delta x^4 \rangle-3\langle \Delta x^2 \rangle^2=0\label{e:kurt}\\
\langle \Delta x^6 \rangle_c\equiv
\langle \Delta x^6 \rangle-10\langle \Delta x^4\rangle\langle\Delta x^2\rangle 
+15\langle \Delta x^2\rangle^3=0\label{e:six}
\end{eqnarray}
respectively.
It is also possible to obtain dimensionless forms of the above expressions
dividing them by the appropriate power of
$\langle\Delta x^2\rangle$.  In this form the fourth cumulant is usually
called the kurtosis, and is a common measure of how close a probability
density is to a Gaussian distribution.

Straightforward manipulations applied to the mean square displacement lead
to the Green-Kubo expression for the diffusion coefficient (if it exists)
\begin{equation}\label{e:GK}
D=\int_0^{\infty}\langle  v_tv_0\rangle dt
\end{equation}
where $v$ is one (say the $x$-) component of the velocity and the subscript
denotes the time.  This
shows that the existence of a diffusion coefficient is also related to
the rate of decay of the velocity autocorrelation function.  The integral
diverges if the decay rate of the integrand is $t^{-1}$ or slower, unless
the integrand is oscillatory.  An oscillatory integrand cannot be ruled out,
but is not typical for the Lorentz gas as observed in numerical simulations.

The diffusion equation is the lowest order macroscopic description of the
diffusion process.  The next approximation, in the direction of 
microscopic space and time scales, {\em ie.} when the probability
density $P$ varies sufficiently rapidly, the right hand side of
Eq.~(\ref{e:diff}) may be augmented by terms such as $B\nabla^4 P$, where
$B$ is called the linear super-Burnett coefficient~\cite{vB}.  Like
the diffusion coefficient, it obeys an Einstein relation
\begin{equation}\label{e:Burn}
\langle \Delta x^4 \rangle-3\langle \Delta x^2 \rangle^2=24Bt
\end{equation}
This is consistent with the zero kurtosis found previously~(\ref{e:kurt}):
to find the kurtosis we divided by $\langle x^2 \rangle^2$ which is
proportional to $t^2$ if the diffusion coefficient exists, while the
Burnett coefficient in~(\ref{e:Burn}) gives a smaller term (for large $t$),
proportional to $t$.  The Burnett coefficient can also be written in the
form of a Green-Kubo relation
in terms of integrals over four-time velocity correlation functions,
\begin{equation}\label{e:GKB}
B=\frac{1}{6}\int_0^{\infty}\int_0^{\infty}\int_0^{\infty}
\left(\langle v_0v_1v_2v_3 \rangle
-\langle v_0v_1\rangle \langle v_2v_3 \rangle
-\langle v_0v_2 \rangle\langle v_1v_3 \rangle
-\langle v_0v_3 \rangle\langle v_1v_2 \rangle\right)
dt_1dt_2dt_3
\end{equation}
where $v_0=v(t=0)$ as above and $v_1=v(t=t_1)$ etc.  A slow decay of the
relevant correlation functions leads more easily to a divergence in this
case than in Eq.~(\ref{e:GK})~\cite{vB}.  This
means that it is possible to have a well defined diffusion coefficient, but
a divergent Burnett coefficient, which occurs if the fourth cumulant
increases faster than $t$ but slower than $t^2$.

\begin{figure}
\begin{picture}(500,530)(-100,0)
\put(-100,30){\epsfig{file=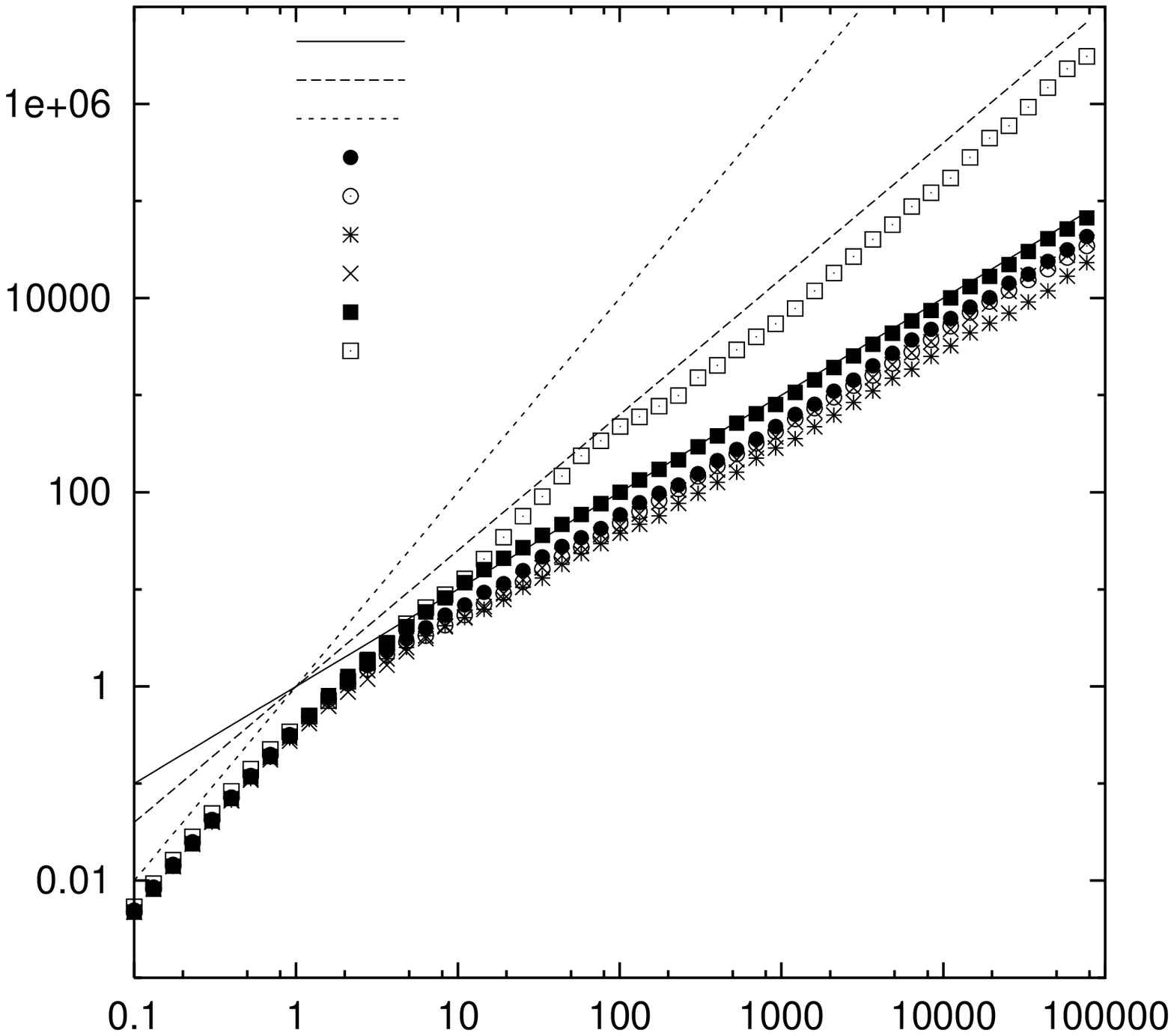, width=500pt}}
\put(170,65){\epsfig{file=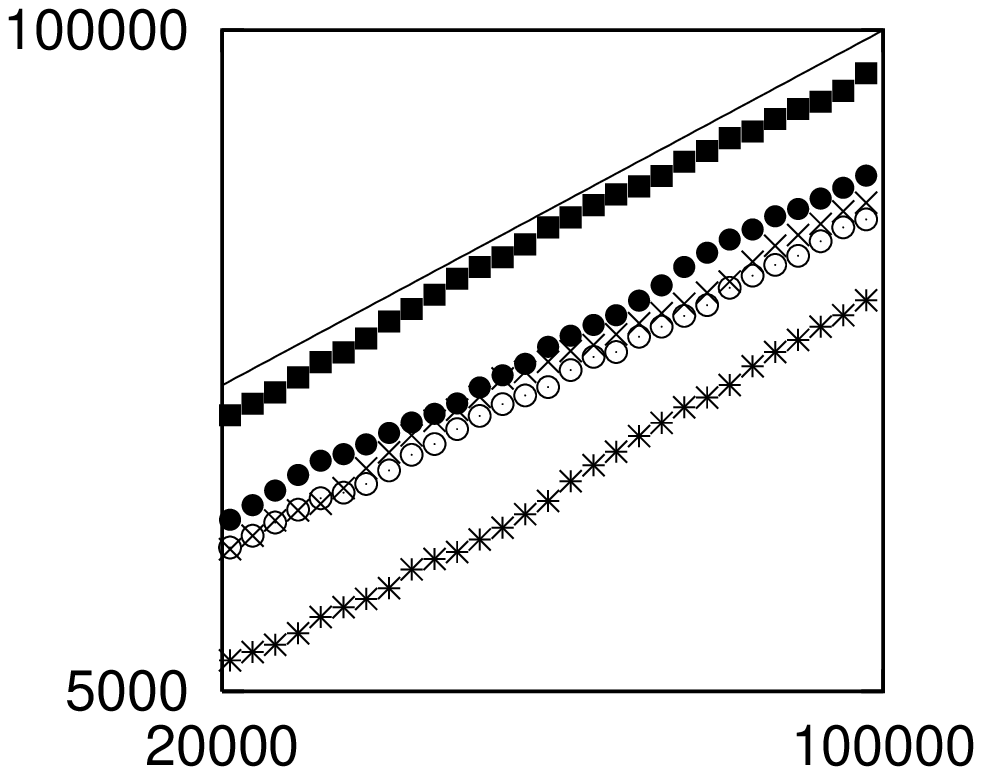, width=200pt}}
\put(155,20){\Large $t$}
\put(-115,265){\Large $\langle\Delta x^2\rangle$}
{\large
\put(0,442){$t$}
\put(0,426){$t^{1.4}$}
\put(0,410){$t^2$}
\put(0,394){L$\infty$}
\put(0,378){LP4}
\put(0,362){R$\infty$}
\put(0,346){RP4}
\put(0,330){F$\infty$}
\put(0,314){FP4}}
\end{picture}
\caption{The mean square displacement (\protect\ref{e:msd}) for the six models
[F,R,L][$\infty$,P4].  In each case there is a transition between ballistic
behavior $\langle \Delta x^2\rangle\sim t^2$ to the long time diffusive (or
superdiffusive) behavior.  The FP4 model is superdiffusive, with an
approximate $t^{1.4}$ law; the others exhibit normal diffusion, $\sim t$.
The diffusion coefficient is largest for the F$\infty$ model, followed
by L$\infty$, RP4, LP4, and R$\infty$ (inset and Tab.~\protect\ref{t:diff});
see the discussion in the text.
\label{f:msd}}
\end{figure}

\begin{figure}
\begin{picture}(300,320)(-70,0)
\put(0,20){\epsfig{file=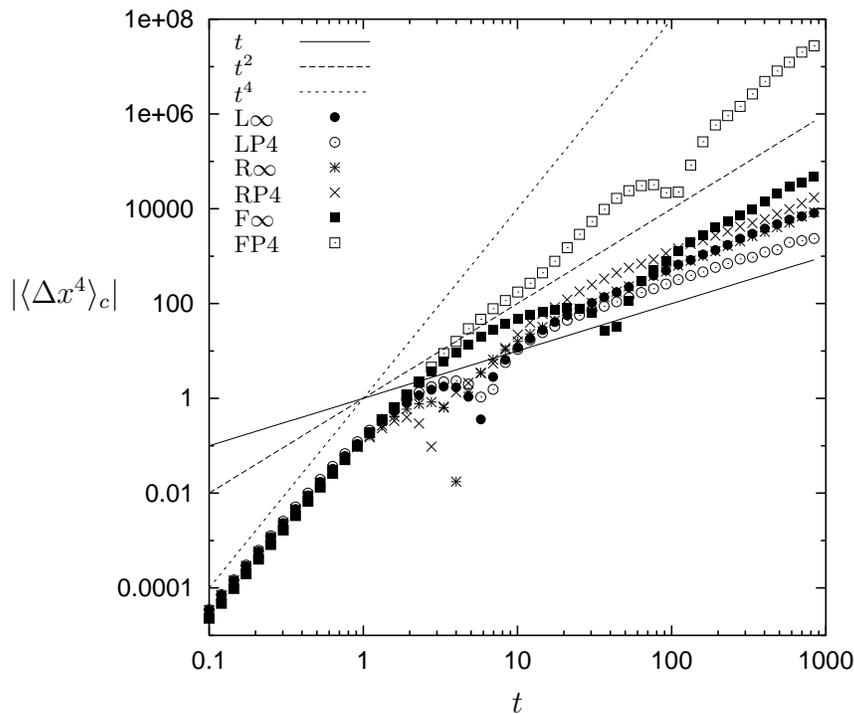, width=300pt}}
\put(165,10){\large $t$}
\put(-25,165){\large $|\langle\Delta x^4\rangle_c|$}
{\small
\put(60,261){$t$}
\put(60,251){$t^2$}
\put(60,241){$t^4$}
\put(60,232){L$\infty$}
\put(60,222){LP4}
\put(60,213){R$\infty$}
\put(60,203){RP4}
\put(60,194){F$\infty$}
\put(60,184){FP4}}
\end{picture}
\caption{The fourth cumulant (\protect\ref{e:kurt}) for the six models
[F,R,L][$\infty$,P4].  We are mostly interested in the long time
behavior, which is Gaussian when the fourth cumulant grows more slowly
than $t^2$,
as holds in all models except the superdiffusive FP4 model.  The periodic
Lorentz gas LP4 has a finite Burnett coefficient, so its fourth cumulant
is proportional to $t$ as shown.  At short times for all models a ballistic
$t^4$ behavior is followed by an intermediate regime in which all models
exhibit a change of sign which appears as a dip due to the logarithmic axes.
\label{f:kurt}}
\end{figure}

\begin{figure}
\begin{picture}(300,320)(-70,0)
\put(0,20){\epsfig{file=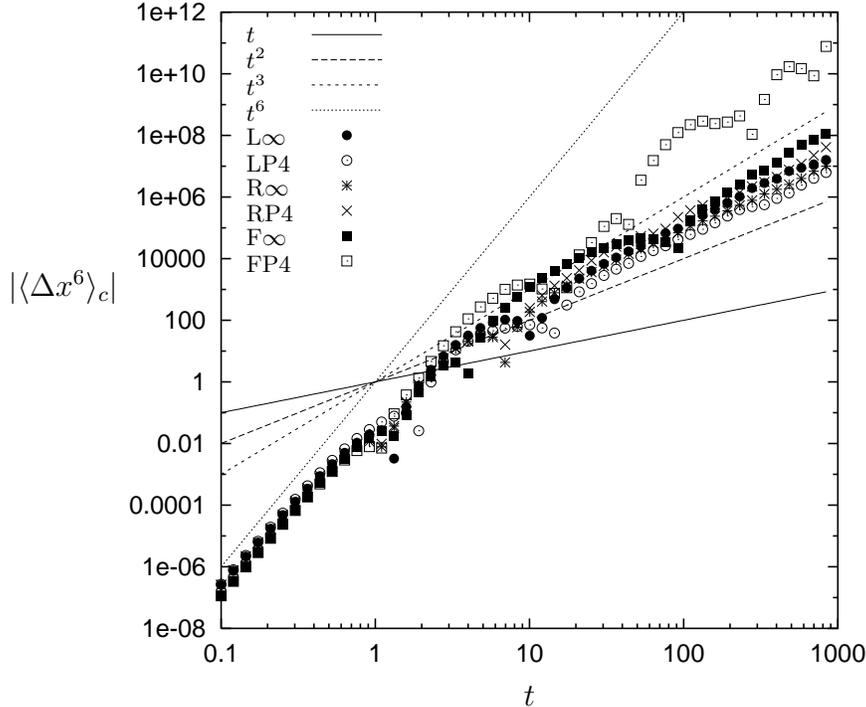, width=300pt}}
\put(165,10){\large $t$}
\put(-30,165){\large $|\langle\Delta x^6\rangle_c|$}
{\small
\put(60,261){$t$}
\put(60,251){$t^2$}
\put(60,241){$t^3$}
\put(60,231){$t^6$}
\put(60,222){L$\infty$}
\put(60,212){LP4}
\put(60,203){R$\infty$}
\put(60,193){RP4}
\put(60,184){F$\infty$}
\put(60,174){FP4}}
\end{picture}
\caption{The sixth cumulant (\protect\ref{e:six}) for the six models
[F,R,L][$\infty$,P4].  The statistics are rather poor in this case, so
that delicate cancelations cannot be observed; these curves should be viewed
as upper limits on the true values for large times.  It is at least
plausible that all of the models except FP4 are Gaussian, which is indicated
by a growth slower than $t^3$.  At small times the ballistic regime gives $t^6$.
\label{f:six}}
\end{figure}

We estimate the cumulants defined in Eqs.~(\ref{e:msd}-\ref{e:six})
numerically by choosing a large number (typically $10^5$)
of random initial conditions for the particle not inside a scatterer, for
a single configuration of scatterers.  We report only the cumulants in
the $x$-direction, although we have checked that the $y$ distribution is
similar.  See Figs.~\ref{f:msd}-\ref{f:six} and Table~\ref{t:diff}.

\begin{center}
\begin{minipage}{6in}
\begin{table}
\begin{tabular}{cccc}
model&F&R&L\\
\hline
$\infty$&$D=0.44$&$D=0.14$&$D=0.27$\\
P4&-&$D=0.24$&$D=0.23,\;\;B=0.12$
\end{tabular}
\caption{Diffusive properties of our models, in terms of the
diffusion coefficient $D$, and the Burnett coefficient $B$ if
they exist.  The existence of these coefficients is related to the
rate of decay of the velocity autocorrelation functions
of~(\protect\ref{e:GK},\protect\ref{e:GKB}).  Note that by this criterion,
the periodic Lorentz gas LP4 is more ``chaotic'' than the infinite Lorentz
model L$\infty$, while the periodic fixed orientation wind-tree
model FP4 is less ``chaotic'' than the infinite model F$\infty$.
Only typical values of the coefficients for the periodic models
are given, since they depend on the position of the scatterers.
\label{t:diff}}
\end{table}
\end{minipage}
\end{center}

For five models, [F,R,L]$\infty$, [R,L]P4, the mean square displacement
is found to be proportional to the time, and the fourth and sixth cumulants
increase slower than $t^2$ and $t^3$ respectively at large times, indicating
that the distribution approaches a Gaussian.  The exception to this rule
is FP4, for which the mean square displacement grows faster than the time,
approximately as $t^{1.4}$ and the kurtosis is also nonzero.  This
superdiffusive behavior is not particularly stable in the sense that the
mean square displacement for this model
has larger fluctuations than the other models, and
the exponent of approximately $1.4$ varies unpredictably between 1 and 2
with the exact positions of the scatterers and the size of the cell
(for example for FP6 or FP8).

\subsection{Connections between chaotic and diffusive properties}\label{s:conn}
We now look at how the chaotic and diffusive properties of our models are
related in the light of our results, specifically of:
(a) normal diffusion without
chaos, (b) superdiffusion without chaos, (c) a possible relation between
the diffusion coefficient and the periodic orbits, and (d) the Burnett
coefficient of the periodic Lorentz gas.  In no way does the data presented
in Figs.~\ref{f:msd}-\ref{f:six} distinguish between chaotic and nonchaotic
models.

\paragraph*{Normal diffusion}
Our results show that it is possible to have a well defined diffusion
coefficient and a Gaussian distribution function without positive Lyapunov
exponents, that is, in the F$\infty$, R$\infty$ and RP4 models.  The RP4 model
is particularly interesting in this regard, since the two most obvious
sources of unpredictability: dispersive collisions and collisions with new
randomly placed scatterers, are absent.  Recall that we argued (in
section~\ref{s:chaos}) that these two processes might together determine
the KS entropy, that is, the rate of loss of information about the particle's
position and velocity.  The RP4 model is thus the clearest example of a
model in which there is normal diffusion, and zero KS entropy.
An example with the same properties, using a periodic rhombus model has
been studied in Ref.~\cite{LRB}, where a mean square displacement proportional
to the time was observed, but the higher moments were not studied.

\paragraph*{Superdiffusion}
The FP4 model, in contrast, has a mean square displacement growing faster
than linearly with the time, and a non-Gaussian distribution.  This can
be related to a slow decay of the velocity autocorrelation function (see above),
but a physical explanation of why this occurs in the FP4 model but not in the
RP4 model is lacking.

\paragraph*{Diffusion and periodic orbits}
It is perhaps somewhat hazardous to deduce more from this data
than the existence of normal diffusion; however
we would nevertheless like to point out
a possible connection between the value of the diffusion coefficient (given
by the vertical displacement of the curves in the inset of
Fig.~\ref{f:msd}) in the
infinite models and the properties of periodic orbits discussed in
section~\ref{s:ISR}.  We make no statement about the diffusion coefficient
of the periodic models because this coefficient depends on the position of the
scatterers.  The idea is
that the presence of periodic orbits in a model with randomly
positioned scatterers might be expected to lower the diffusion coefficient,
since the periodic orbits may keep the particle trapped in roughly the
same place; this takes time without contributing to the mean square
displacement.  Thus we observe that the fixed orientation wind-tree model
F$\infty$ has no periodic orbits and has the largest diffusion coefficient;
the random Lorentz gas L$\infty$ has periodic orbits, but they are exponentially
unstable, and has the next largest diffusion coefficient; the random
orientation wind-tree model R$\infty$ has periodic orbits with power law
instability that can inhibit diffusion for a long time,
and the diffusion coefficient is the smallest of the three.

\paragraph*{The Burnett coefficient}
The fourth cumulant grows linearly with time for the LP4 model, indicating
a finite Burnett coefficient, as expected, given its exponential decay of
correlations~\cite{CD}.  The other models, F$\infty$, R$\infty$, RP4 and
L$\infty$ have a divergent Burnett coefficient, which is not surprising
since their decay of correlations is most likely as a power law.  
Thus, from the more subtle properties of the two time (displacement)
correlation function, we can deduce some properties relating to the rate
of decay of higher order correlations, even special four-time (velocity)
correlations,
but not whether there is a positive Lyapunov exponent, since L$\infty$ behaves
similarly to R$\infty$ from this point of view.  Even in the intermediate
region between the ballistic behavior at small times and the (super)-diffusive
behavior at large times in Fig.~\ref{f:msd}, that is, where the distribution
functions are measured at microscopic time scales, there is no clear
distinction between the chaotic and nonchaotic models.

We have now discussed at some length the connection between chaotic and
diffusive properties of our models.  While this has yielded valuable
insight into how little chaoticity is needed for diffusion to occur,
it has not provided us with an understanding of how chaotic and
nonchaotic diffusion differ, let alone a method to distinguish chaotic
and nonchaotic diffusive systems experimentally.  The reason, given in
Ref.~\cite{G97} is that two time correlations (for example our cumulants)
are insufficient to characterize chaos; multi-time correlations (or
their underlying probability distributions) are required.  The remaining
sections of this paper all use multi-time distributions of one form or
another to investigate our models.  We begin
with a more detailed study of the Grassberger-Procaccia method than
in Ref.~\cite{DCvB}, where we found it to incorrectly classify the F$\infty$
model as chaotic.

\section{The Grassberger-Procaccia method}\label{s:GP}
Grassberger and Procaccia~\cite{GP} gave the earliest practical methods
for computing chaotic properties such as entropies and dimensions
from experimental or computer generated time series. These are
still the most popular methods in use, sometimes with minor
variations.  In Ref.~\cite{GBFSGDC} the conclusion of
microscopic chaos using data from a Brownian motion experiment was based
on a slightly different method of Cohen and Procaccia~\cite{CP}.
Both methods are clearly reviewed in Ref.~\cite{ER}. 

We consider these methods as applied to the calculation of the
KS entropy $K$, the non-zero value of which characterizes
chaos.  We recall the discussion of section~\ref{s:chaos}
where we concluded that the wind-tree models all have zero KS entropy,
but the Lorentz gas has a KS entropy equal to its positive Lyapunov exponent.

The original Grassberger-Procaccia method~\cite{GP} computes a slightly
different dynamical entropy $K_2$ (defined using the square of the probability,
$p^2$ rather than the conventional $p\log p$) satisfying the inequality $K_2<K$.
Thus a positive estimate $K_2>0$ implies that $K>0$ and hence chaoticity.
We follow Ref.~\cite{GBFSGDC} in using an adaptation due
to Cohen and Procaccia~\cite{CP} that allows $K$ to be estimated directly.

We replace the experimental time series of the positions of the Brownian
particle in Ref.~\cite{GBFSGDC} by a numerical time series of any of our
infinite or periodic diffusive models, containing the $x$ and $y$
positions of the particle at $10^6$ times uniformly spaced by a separation
$\Delta t$.  Unlike Ref.~\cite{GBFSGDC},
\footnote{Because there is only
one experimental time series in Ref.~\cite{GBFSGDC}, the authors generated
new time series with time steps equal to a multiple of $\Delta t$ by removing
data points.  The new time step was denoted $\tau$.  We do not need to do
this in our numerical work, and we do not use the notation $\tau$ for time
steps to avoid confusion with the mean free time.}
our values of $\Delta t$, equal to
1 and 0.01 time units are close to the dynamical time scale determined by the
mean free time $\bar\tau$ (see Sec.~\ref{s:time}). 

A single point on our numerical
trajectory is denoted $(x_i,y_i)$, where $i$ is an integer ranging from 0
to $10^6-1$. For the fixed oriented square (F) models the trajectory contains
all the phase space information except the velocity which can take only
four values in this case.  For the randomly oriented square (R) and the
Lorentz (L) models the
velocity has one continuous degree of freedom.  The Brownian motion
experiment uses only one component (say, $x$) of the particle position,
ignoring the huge number of other degrees of freedom contained in the fluid
system.
From a mathematical point of view, as long as the degrees of freedom are all
coupled (that is, all the degrees of freedom depend on each other directly
or indirectly), this type of method usually converges to the correct entropy
(or dimension) using as few as one measured variable, however the
efficiency may then well be such that an unreasonably large number of data
points is required for a reliable estimate of these quantities. 

A number (typically $10^2$ to $10^3$) of (in our case non-overlapping)
uniformly spaced subsequences of the long trajectory are denoted
``reference segments'' (``reference trajectories'' in Ref.~\cite{GBFSGDC}),
having lengths $n$ ranging from 1 to 100.  These reference segments are
compared with all subsequences of the trajectory having the same length.
The comparison is made according to the position relative to the initial
point of each subsequence,
with a Euclidean metric in real space and a
maximum metric over the trajectory.  Thus, if $i$ and $j$ give the positions
in the full trajectory of the initial points of the $M$ reference segments
and $N\approx 10^6$ total segments respectively, the distance between
the segments beginning at $i$ and $j$ is defined as
\begin{equation}\label{e:dist}
d_n(i,j)=\max_{k=0}^{n-1}\sqrt{(\Delta x_{i+k}-\Delta x_{j+k})^2
+(\Delta y_{i+k}-\Delta y_{j+k})^2}
\end{equation}
where $\Delta x_{i+k}=x_{i+k}-x_i$ etc. 

We are now in a position to count the number of recurrences, in what is
called ``pattern probability'' in~\cite{GBFSGDC}.  For a given tolerance
or spatial resolution $\epsilon$ we can define the probability of a certain
pattern defined by the reference sequence beginning at $i$ as the
proportion of all the subsequences of the full trajectory that are
within a distance (as defined above) $\epsilon$ of the reference sequence:
\begin{equation}
P(i,n,\epsilon,\Delta t)=\frac{1}{N}\#_{j}[d_n(i,j)<\epsilon]
\end{equation}
and from this the ``pattern entropy''
\begin{equation}\label{e:patt}
K(n,\epsilon,\Delta t)=-\frac{1}{M}\sum_i\log_{10}P(i,n,\epsilon,\Delta t)
\end{equation}
which gives the information theoretic entropy of the dynamics with respect
to the pattern (of length $n$ and resolution $\epsilon$) probability
distribution.  The pattern entropy thus contains information about the
$n$-time (displacement) distribution functions, as suggested by the
discussion at the end of the previous section.

When the pattern entropy increases linearly with time in the limit of
large $n$, it is possible to define an entropy per unit time,
\begin{equation}
h(\epsilon,\Delta t)=\frac{1}{\tau}\lim_{n\rightarrow\infty}
[K(n+1,\epsilon,\Delta t)-K(n,\epsilon,\Delta t)]
\end{equation}
In practice $n$ is finite, and $K$ is always bounded above by
$\log_{10}N$ (since it is an average of $-\log_{10}P$, where $P$ bounded
below by $1/N$) and tends to saturate towards this value at long times.
However, we can often find a large enough linear region in a plot of
$K$ as a function of $n$ (for fixed $\epsilon$ and $\Delta t$)
to compute $h(\epsilon,\Delta t)$.  In
Ref.~\cite{GBFSGDC} $h(\epsilon,\Delta t)$ was plotted as a function of
$\epsilon$ with different curves corresponding to different $\Delta t$.
From dimensional arguments it follows that any diffusive process has a scale
invariance leading to a dependence $h\sim 1/\epsilon^2$, which is in fact
observed in such plots.

Assuming that there is sufficient data that all the limits (of large $N$, $M$
and $n$) can be approximated reliably, the KS entropy $K$ is
equal to the maximum value of $h(\epsilon,\Delta t)$ as $\epsilon$ and
$\Delta t$ are varied.  The measured value $h(\epsilon,\Delta t)$ will
be less than the true KS entropy $K$ if the values of $\epsilon$ and or
$\Delta t$ are so large that not all of the information contained
in the dynamics is represented in the time series.  This effect does not
explain why a nonchaotic system may appear to be chaotic, that is, why the
measured value of $h(\epsilon,\Delta t)$ is greater than $K=0$, as in
Ref.~\cite{DCvB}: this must be due to a problem with the above
mentioned limits, and we defer a discussion of this point until after
we have presented our results.
 
\begin{figure}
\begin{picture}(330,530)(-70,0)
\put(0,360){\epsfig{file=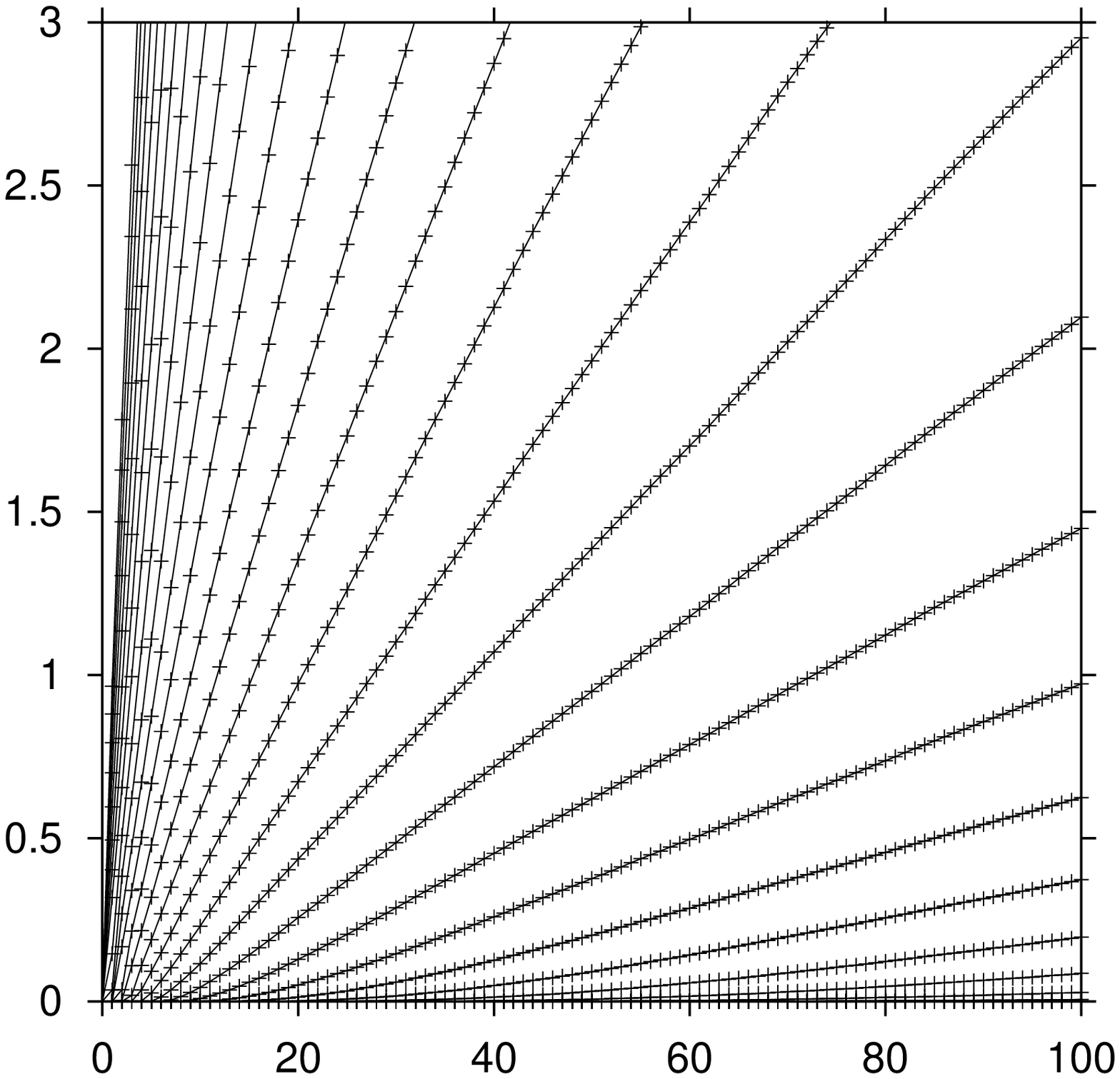, width=160pt}}
\put(170,360){\epsfig{file=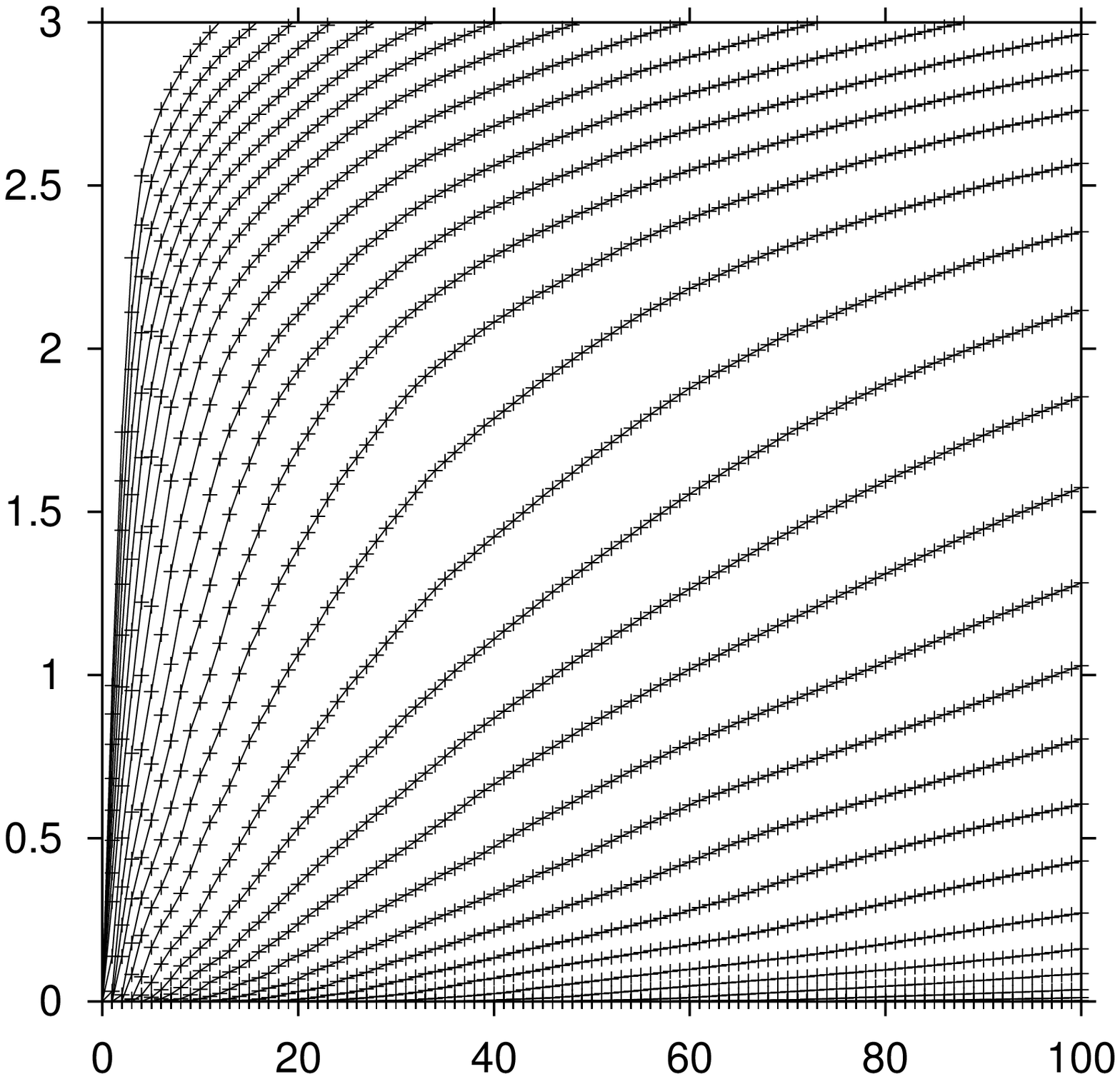, width=160pt}}
\put(0,190){\epsfig{file=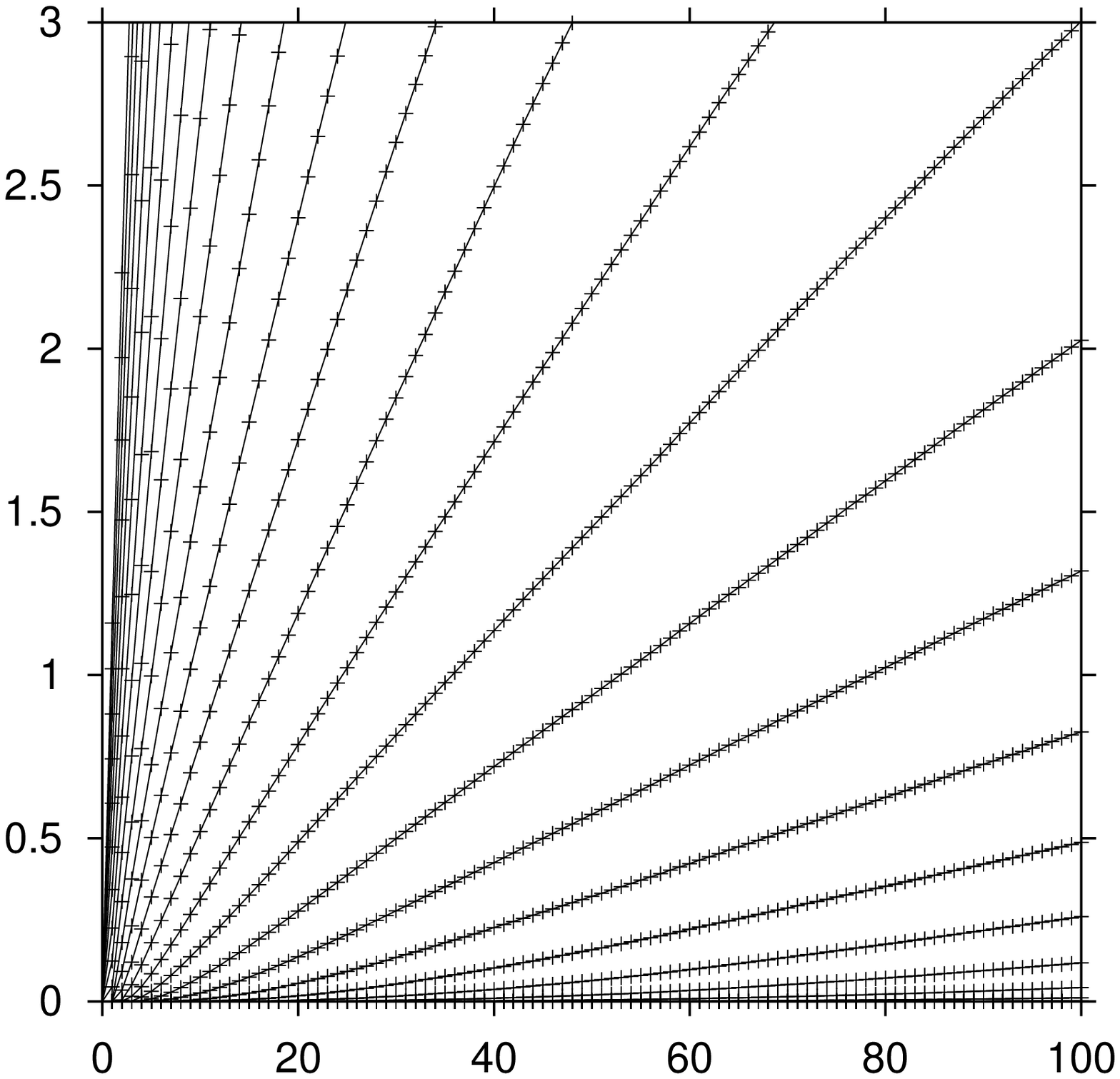, width=160pt}}
\put(170,190){\epsfig{file=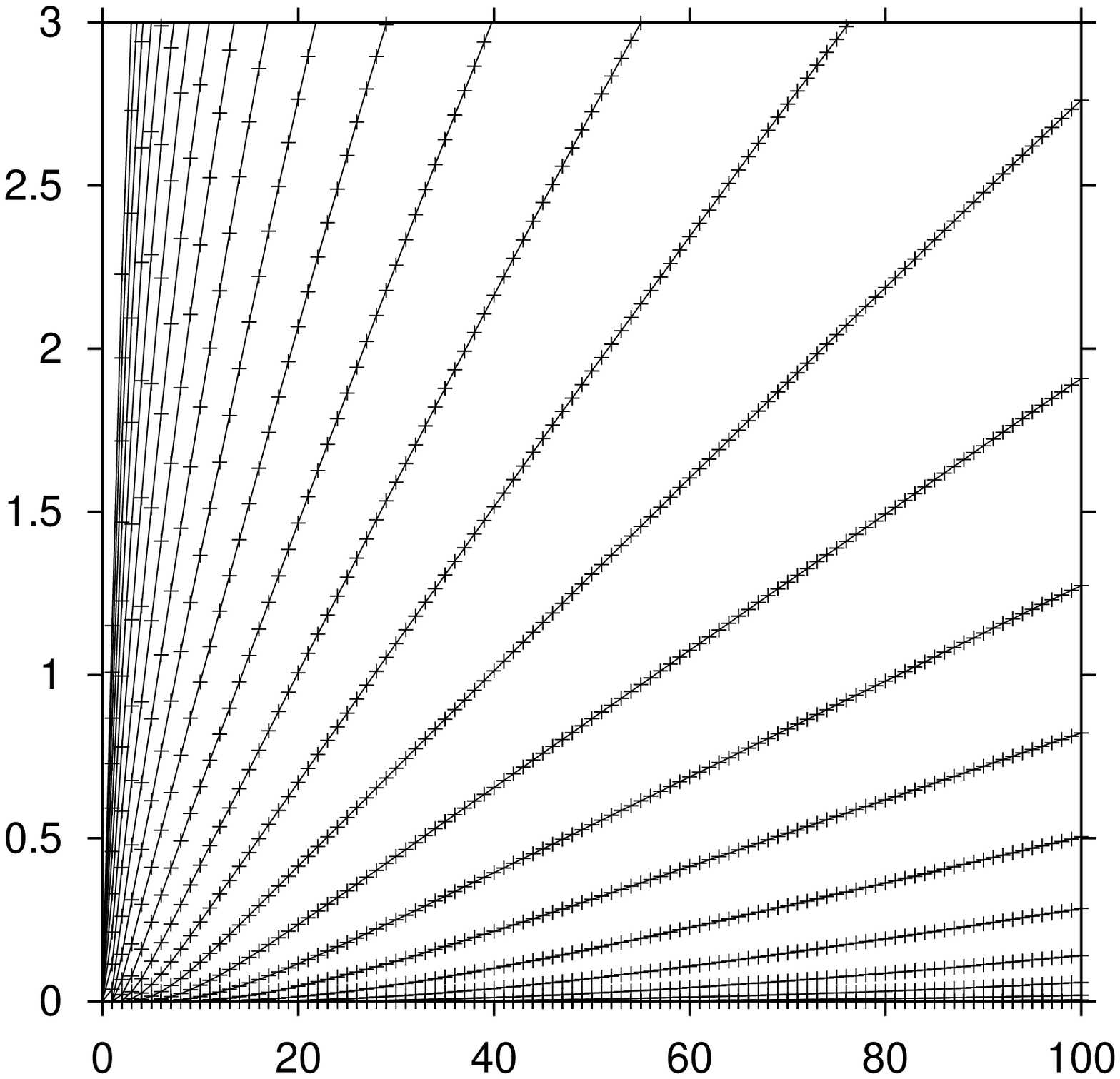, width=160pt}}
\put(0,20){\epsfig{file=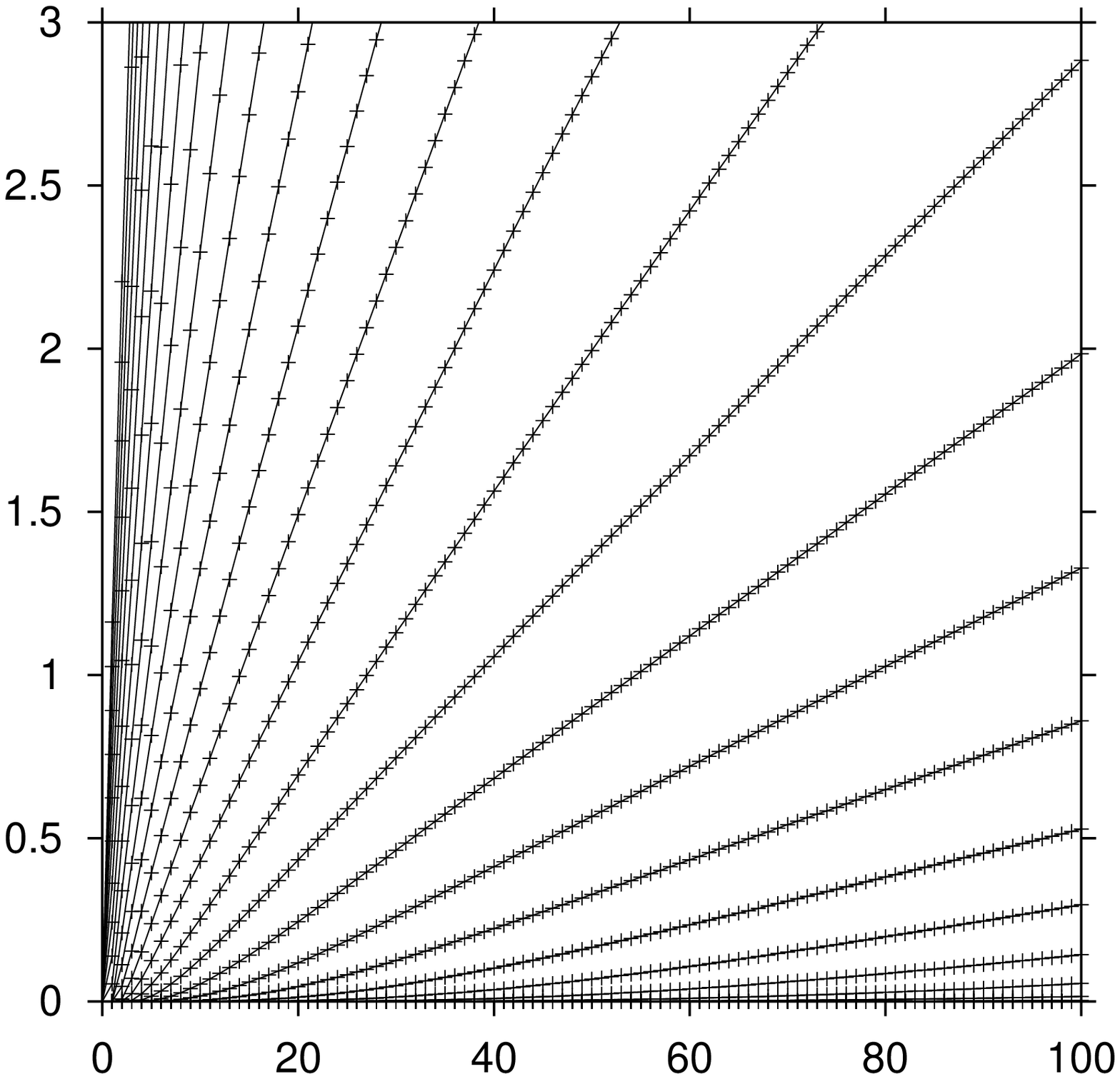, width=160pt}}
\put(170,20){\epsfig{file=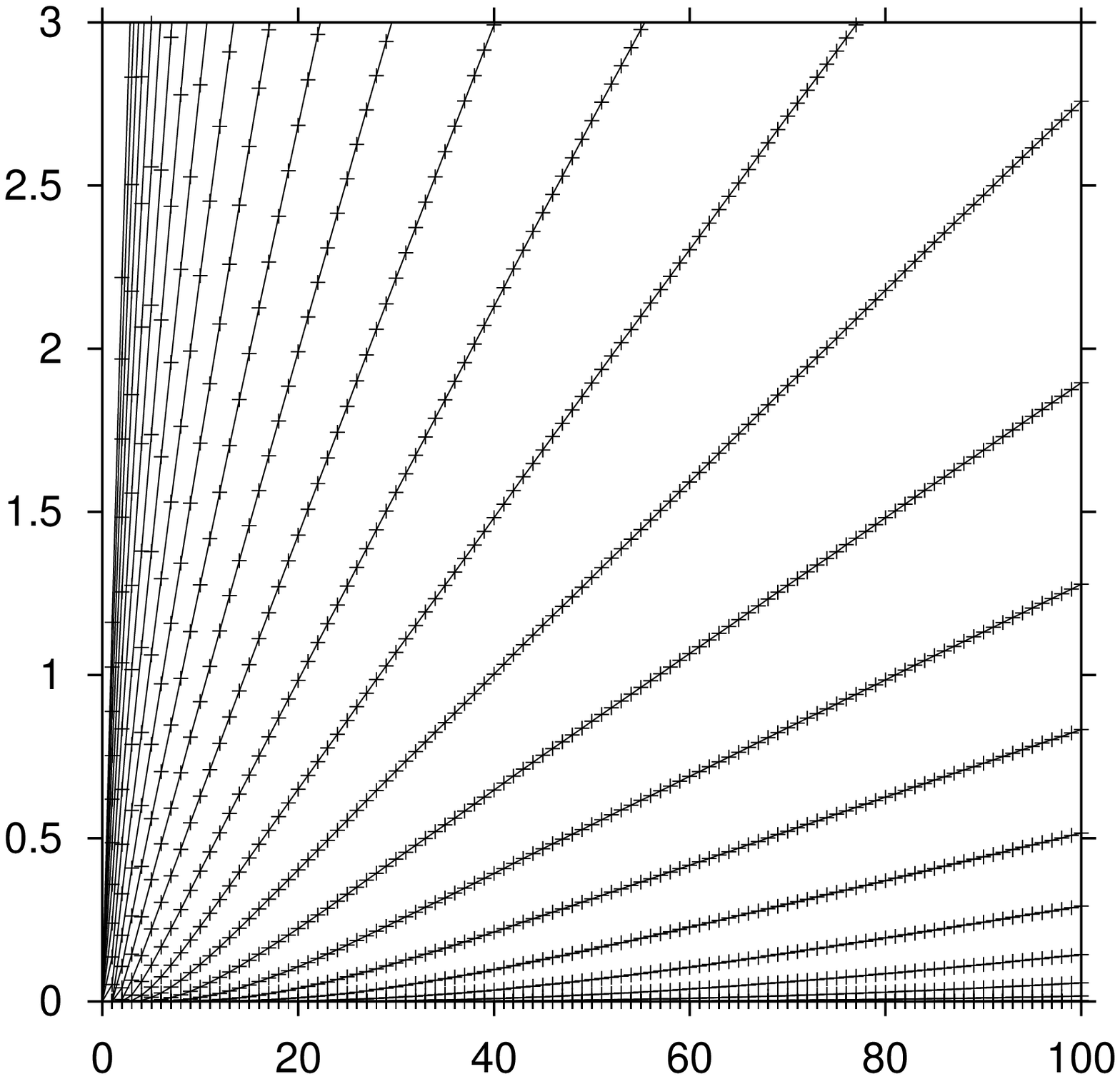, width=160pt}}
\put(85,10){\large $n$}
\put(-65,95){\large $K(n,\epsilon,\Delta t)$}
{\large
\put(-10,480){F}
\put(-10,310){R}
\put(-10,140){L}
\put(85,515){$\infty$}
\put(245,515){P4}}
\end{picture}
\caption{The pattern entropy $K(n,\epsilon,\Delta t)$,
Eq.~(\protect\ref{e:patt}),
plotted against the segment length $n$ for a time step $\tau=1$.  The models
(see Sec.~\protect\ref{s:not}) are F (upper), R (middle) and L (lower), each
with $\infty$ (left) and $P4$ (right).  The different curves on each graph
correspond to different values of $\epsilon$ from approximately $0.3$ to
$300$ in multiples of $1.21$; large values of
$\epsilon$ correspond to smaller pattern entropies since the probability of
recurrences $P(i,n,\epsilon,\tau)$ is then larger.  A linear behavior indicates
positive KS entropy, so both chaotic and nonchaotic models appear to be
chaotic except
the superdiffusive model FP4; see the discussion in the text.\label{f:GPL}}
\end{figure}

\begin{figure}
\begin{picture}(330,530)(-70,0)
\put(0,360){\epsfig{file=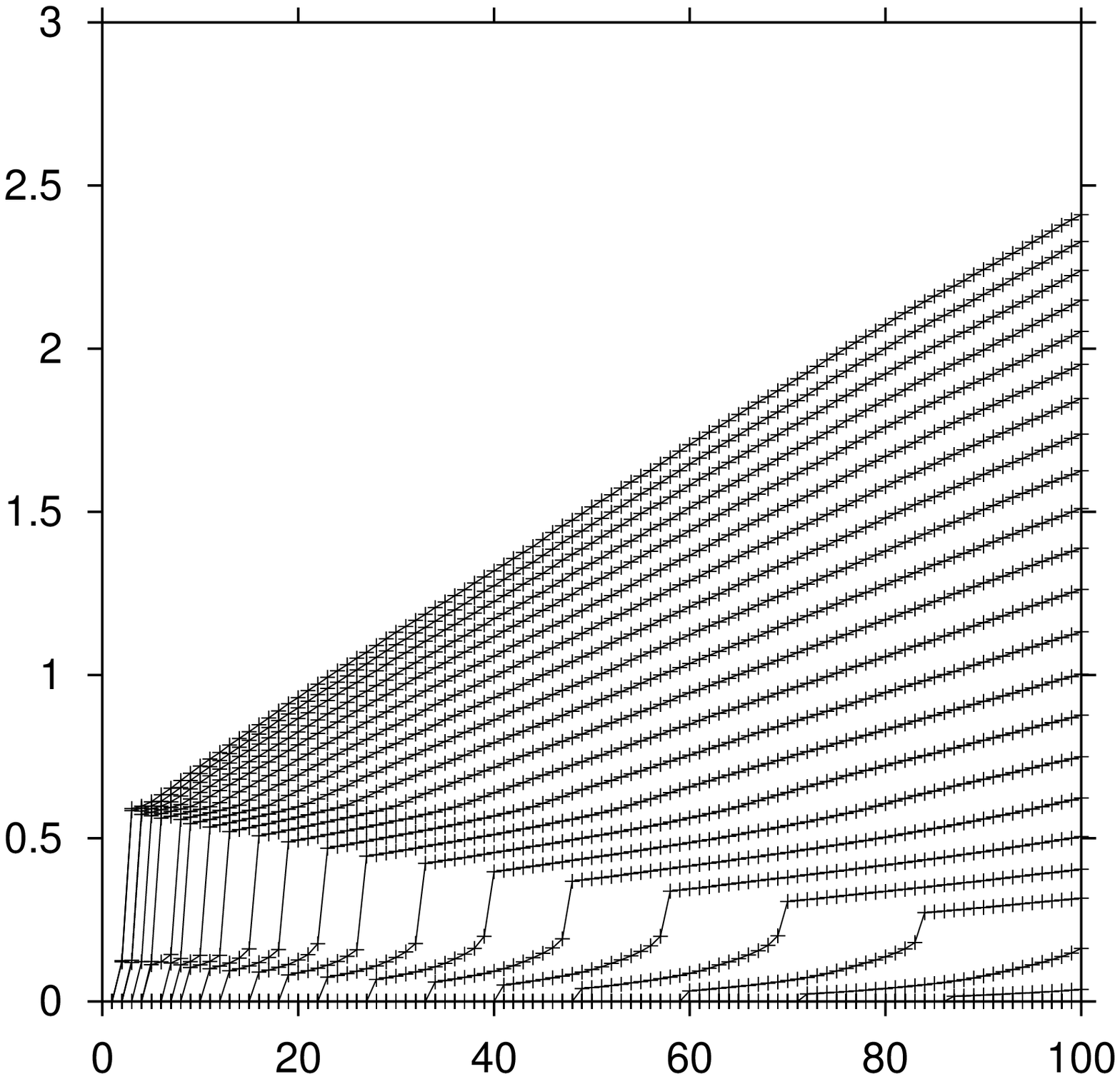, width=160pt}}
\put(170,360){\epsfig{file=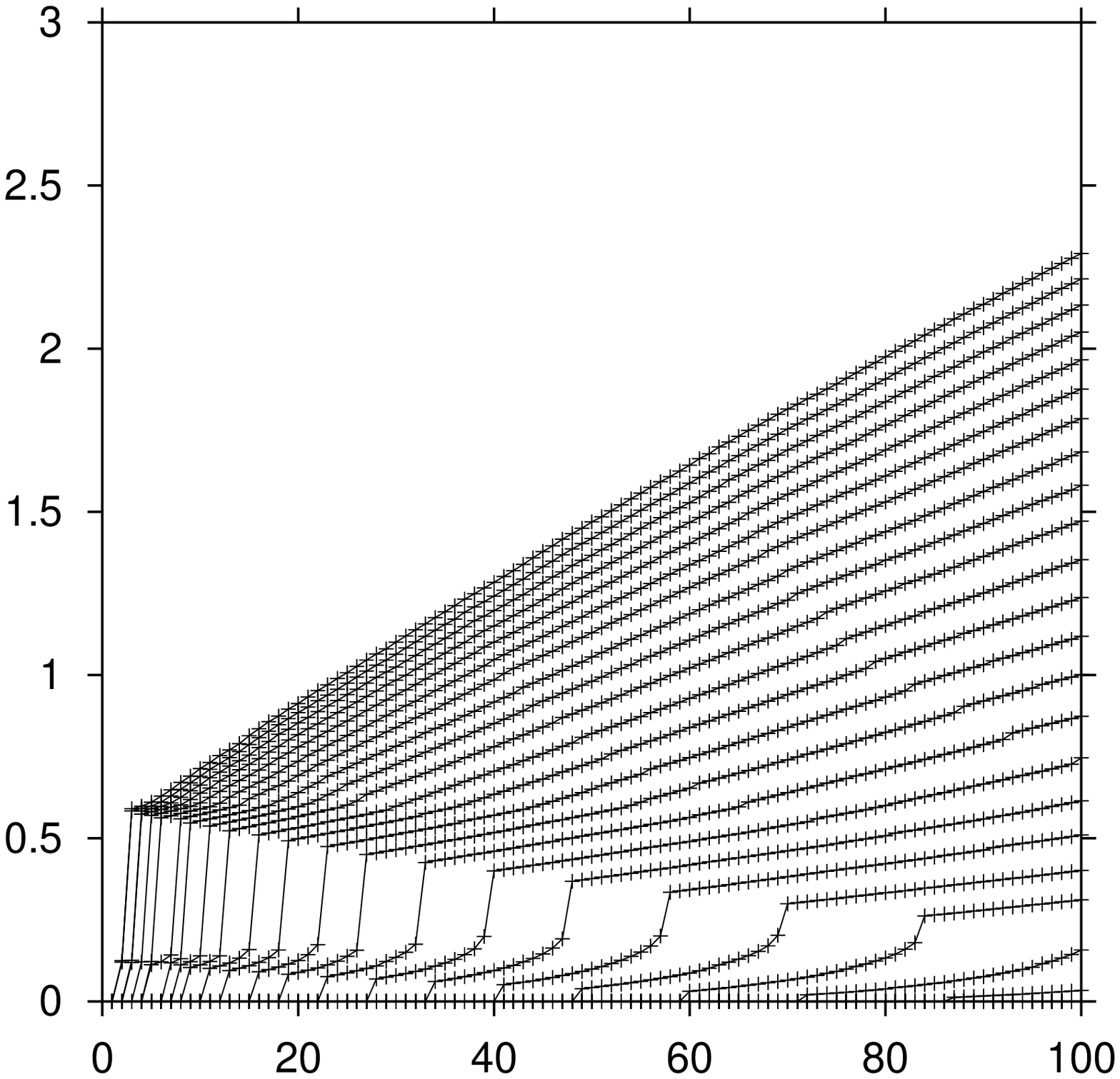, width=160pt}}
\put(0,190){\epsfig{file=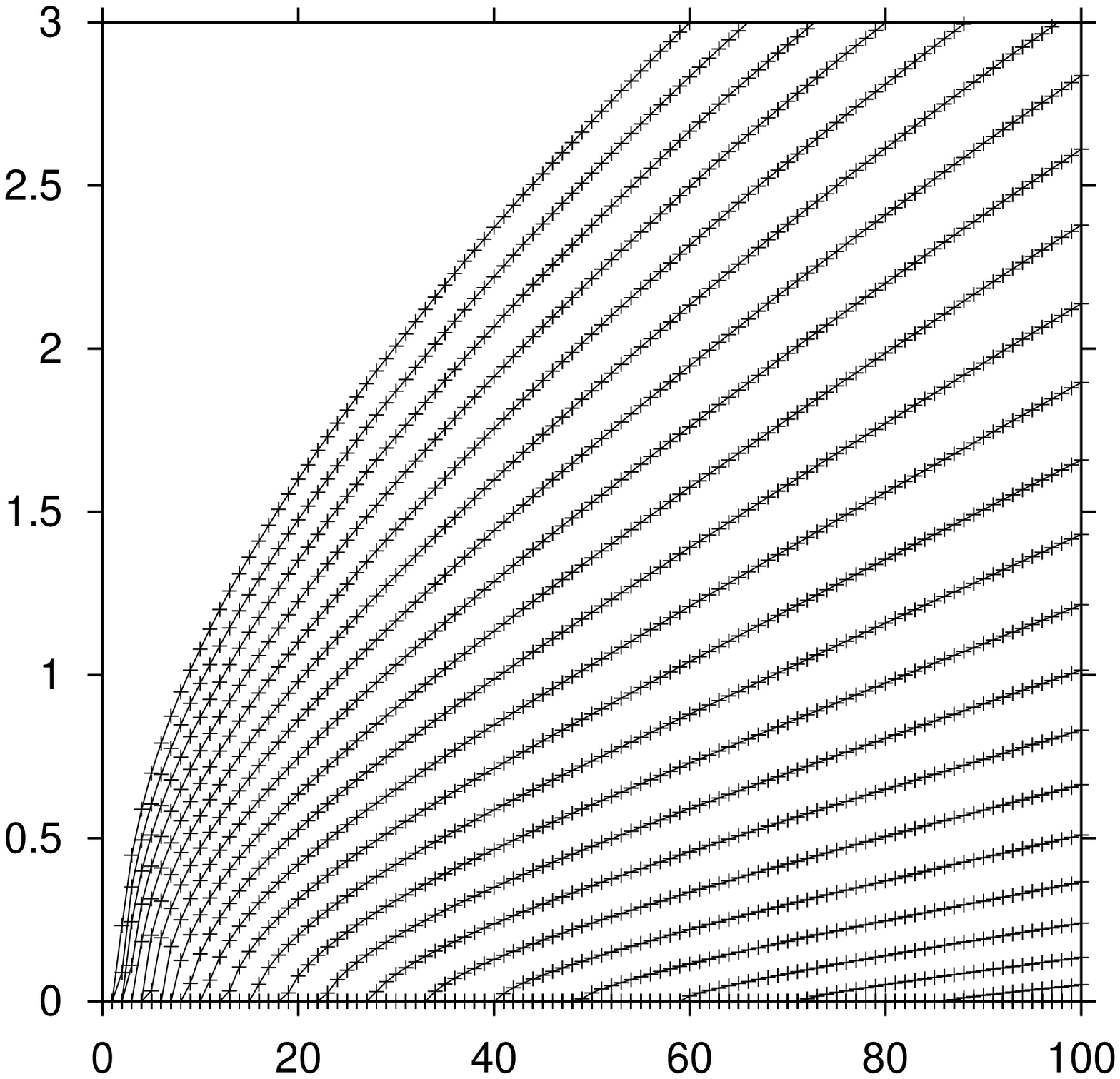, width=160pt}}
\put(170,190){\epsfig{file=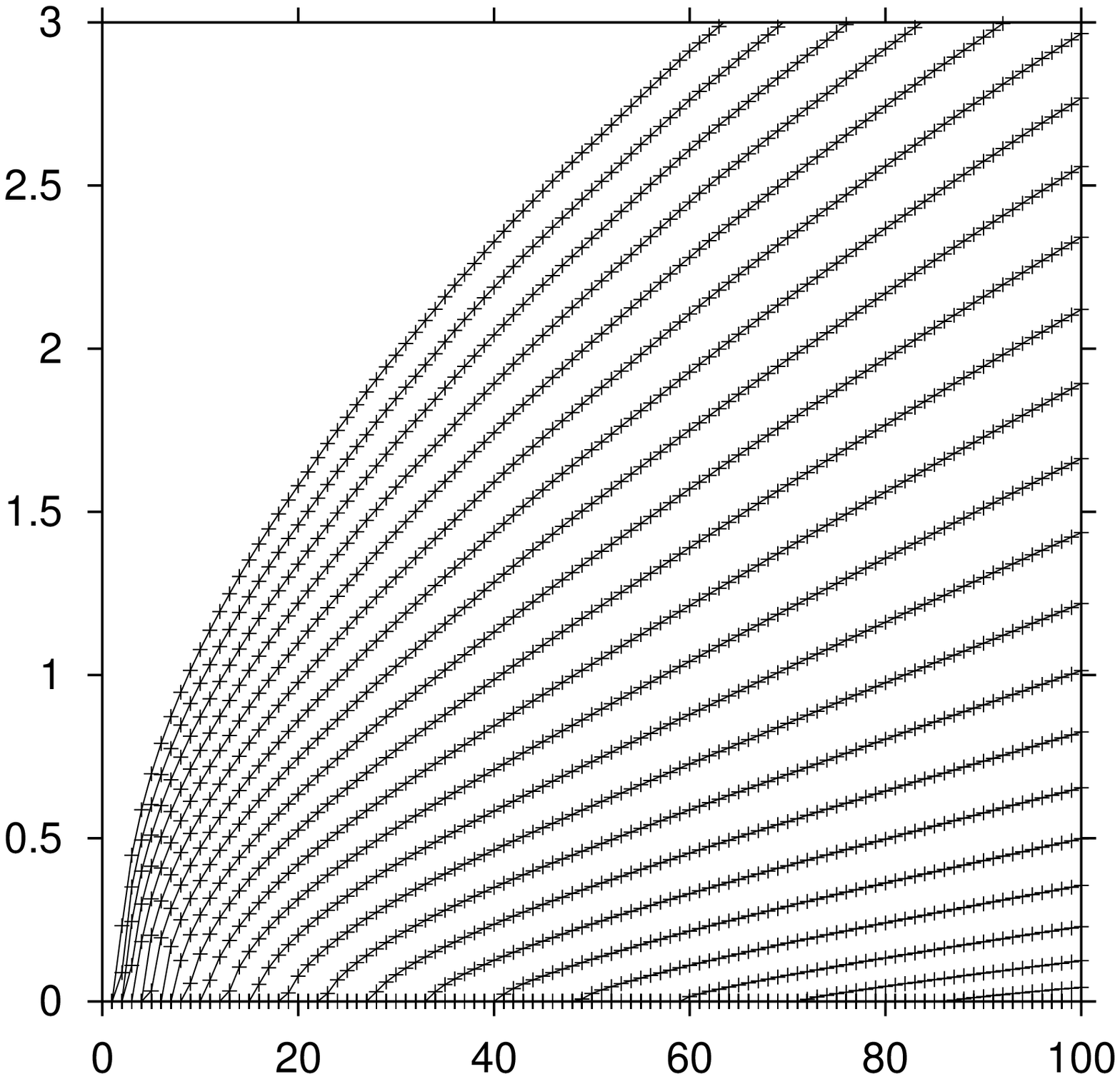, width=160pt}}
\put(0,20){\epsfig{file=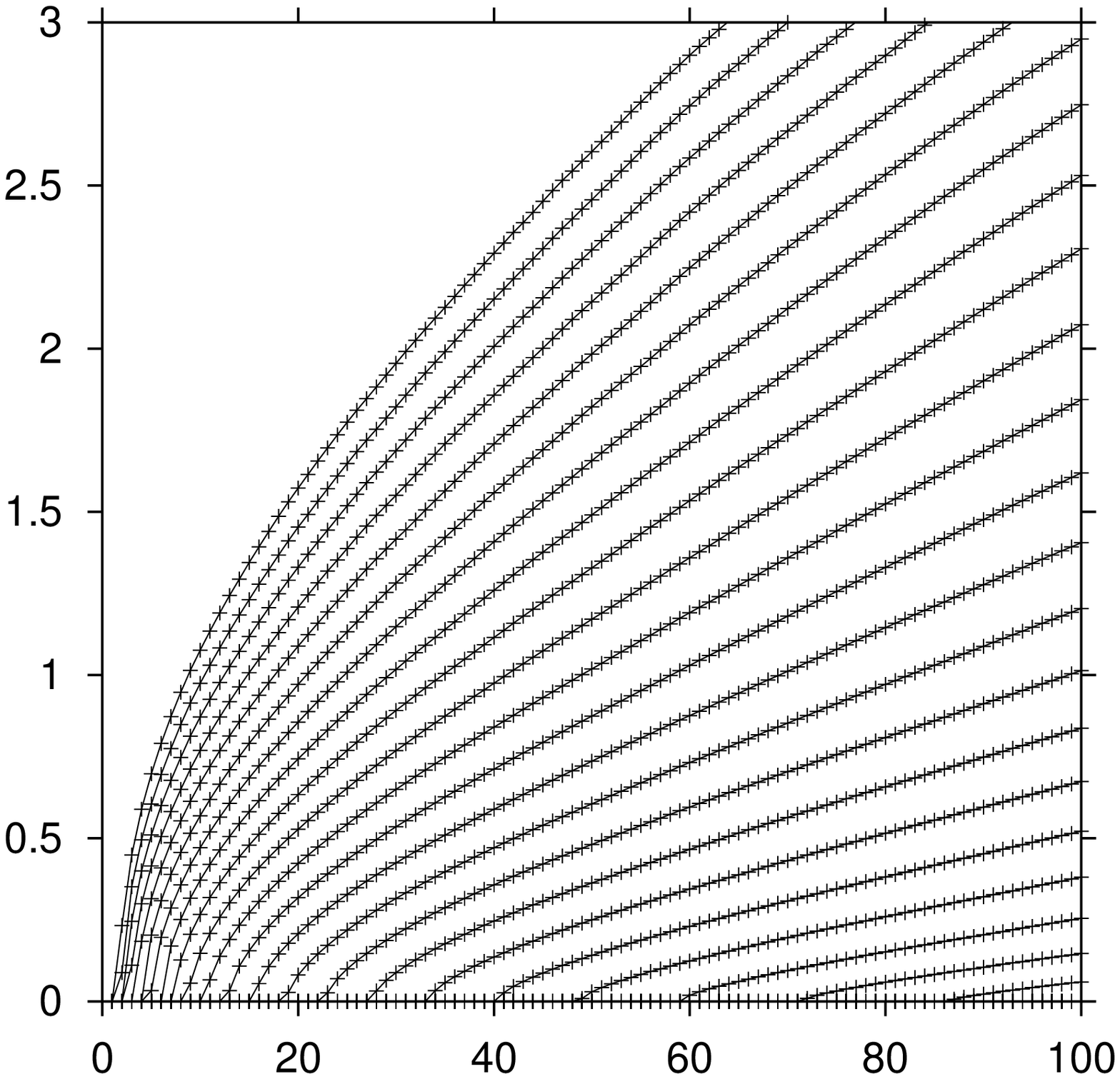, width=160pt}}
\put(170,20){\epsfig{file=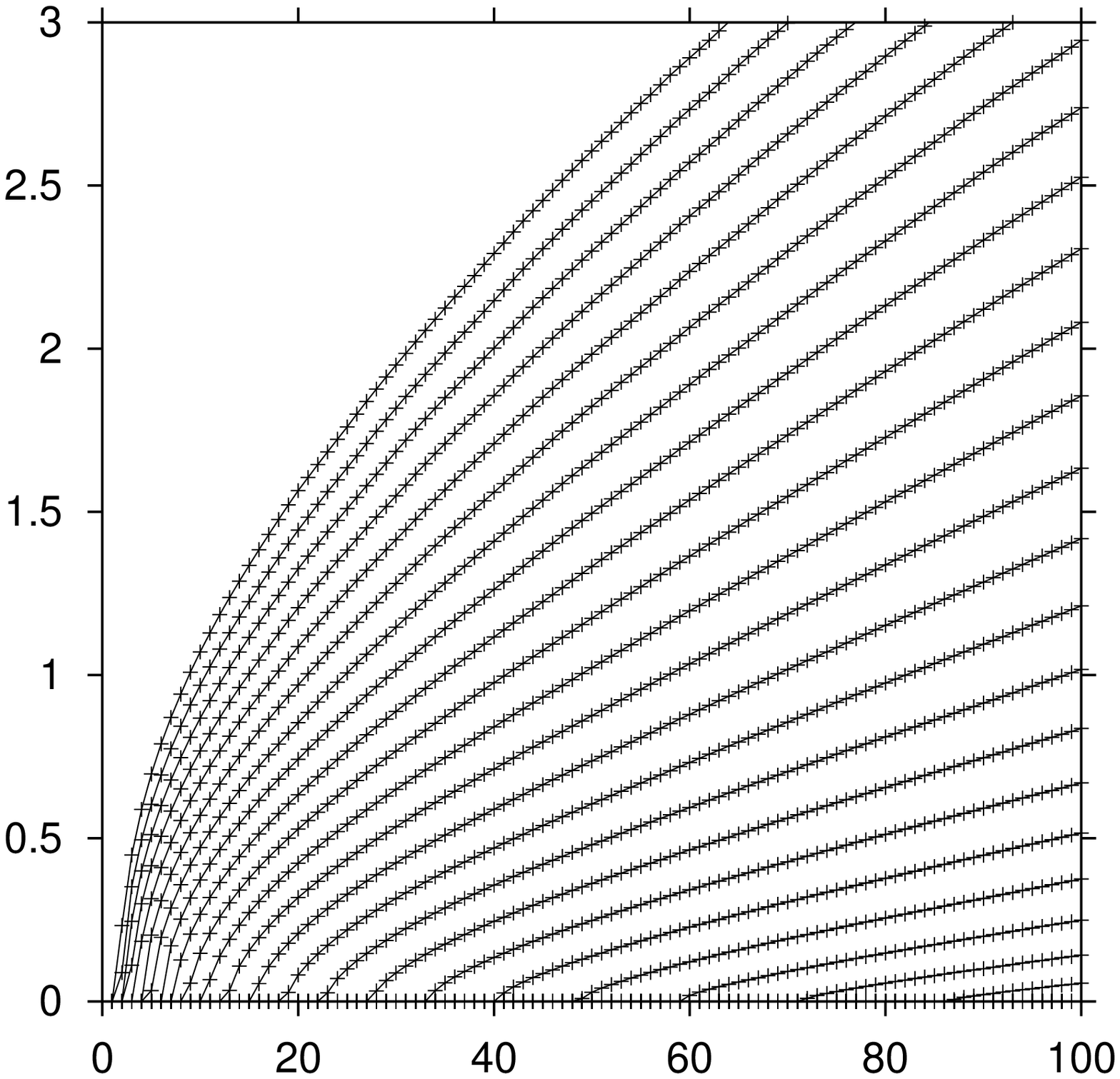, width=160pt}}
\put(85,10){\large $n$}
\put(-65,95){\large $K(n,\epsilon,\Delta t)$}
{\large
\put(-10,480){F}
\put(-10,310){R}
\put(-10,140){L}
\put(85,515){$\infty$}
\put(245,515){P4}}
\end{picture}
\caption{The pattern entropy $K(n,\epsilon,\Delta t)$,
Eq.~(\protect\ref{e:patt}),
plotted against the segment length $n$ for a time step $\Delta t=0.01$.
See Fig.~\protect\ref{f:GPL} for details, except that the range of $\epsilon$
is now approximately $0.03$ to $30$.  At short times, the trajectories are
mostly uninterrupted
straight lines, and hence do not appear chaotic; the lines curve with
decreasing gradient.  At longer times, there are some collisions, leading
to a straightening (more chaotic behavior).  The chaotic and nonchaotic
models look the same, as in
Fig.~\protect\ref{f:GPL} except that the particle in the F models has
only four velocity directions, leading to the characteristic feature below an
entropy equal to $\log_{10}4\approx 0.6$.
\label{f:GPS}}
\end{figure}

Our results for each of the six models [F,R,L][$\infty$,P4] are illustrated
in Fig.~\ref{f:GPL} with $\Delta t=1$ and in Fig.~\ref{f:GPS} with
$\Delta t=0.01$.  The case $\Delta t=1$ samples the motion of the particle as
it begins to diffuse among the scatterers.  The models (both chaotic and
nonchaotic) are indistinguishable, with the linear growth of the pattern
entropy suggesting positive KS entropy, except that the superdiffusive model
FP4 is showing signs that the KS entropy is zero, as a nonchaotic model should. 
 
For $\Delta t=0.01$ (Fig.~\ref{f:GPS}), where mostly ballistic behavior
occurs, the graphs of pattern entropy vs. number of time steps is no longer
linear.  The R and L models look much the same, while the F models
have an unusual feature when the pattern entropy is equal to
$\log_{10}4\approx 0.6$
which is due to the four available wind directions, that is, only a fourth
of the trajectories at a given $(x,y)$ point are close in phase space after a
certain time.  The pattern entropy is becoming linear at larger times in
all models, suggesting positive KS entropy.

We can draw the following conclusions about the use of Grassberger-Procaccia
type methods for this class of problems: (a) measurements need to be
made at microscopic time scales, (b) even at microscopic time scales,
the method does not seem to distinguish between chaotic and nonchaotic
dynamics, and (c) the reason seems to be that unfeasibly long time series
are needed in order to do that.

\paragraph*{Macroscopic measurements}
It is clear that in an analysis
as explained above, using large distance and time scales a diffusive system
shows itself as almost
independent of its microscopic dynamics (whether chaotic, non-chaotic or
stochastic), although we cannot exclude the possibility that an unreasonably
long trajectory might still give some evidence of weak correlations that
survive for macroscopic times. 
In the Brownian motion experiment, for example, there are of the order of
$10^{10}$ collisions of the Brownian particle with the surrounding solvent
between each measurement
of its position, so any correlations arising from, say, nonchaotic microscopic
dynamics would be effectively averaged out between measurements.  This is
essential from the point of view of designing future experiments, but it is
not the full story as it does not explain our results, which are obtained
using short times.
The Brownian motion experiment could not investigate the motion at short time
scales, but it is important to know whether data from an improved experiment
might determine the microscopic chaoticity in principle.

\paragraph*{Microscopic measurements}
Our results do not suggest that the Grassberger-Procaccia type methods
can distinguish between chaotic and nonchaotic diffusive models even at
short times because
the nonchaotic R models consistently give the same results as the chaotic
L models.  The only distinctions we have been able to make are between 
diffusion and superdiffusion for $\Delta t=1$, and between the continuous
velocity space of the R and L models and the four velocity directions of
the F models.  Both of these distinctions can be made from a time series
without using the intensive analysis of the Grassberger-Procaccia method.

\paragraph*{How long a time series is required?}
Our wind-tree models appear unpredictable (in the sense of positive KS entropy)
because it takes the system
some time to ``realize'' that it is nonchaotic, in other words, quite a
long time series length $N$ is required.  For example, the most pronounced
nonchaotic diffusive model we have is RP4.  Once the particle ``knows'' the
positions and orientations of all the scatterers, the system is completely
predictable, and hence a zero KS entropy is manifest.  However it is
necessary for the particle to enumerate the large number of ways of achieving
this%
\footnote{A conservative and very rough estimate for the number of ways
for the particle to determine the positions and orientations of the
scatterers is as follows:  It must collide with tow adjacent sides of all
four scatterers, a total of eight collisions.  For each scatterer there
are four possible pairs of adjacent sides, making $4^4=256$ combinations.
The eight collisions could occur in any order, giving an extra factor
of $8!=40320$.  Thus our estimate is $4^4\times 8!\approx 10^7$.  Note that
collisions may not be able to occur in any order (thus lowering the
estimate), but we have ignored situations where the particle collides with
three or four sides of a scatterer, or with the same sides more than once
(thus greatly increasing the estimate).  The trajectory length would need
to be substantially larger than the true number of ways the particle can
determine the scatterer configuration, in order to obtain reasonable
statistics, so that nonchaotic recurrences dominate.}
with sufficient statistics that nonchaotic recurrences%
\footnote{Nonchaotic recurrences are those with a probability that
decreases more slowly than exponentially with length, indicating greater
probability than is characteristic of chaotic systems, and hence zero
KS entropy.}
dominate the GP calculation, for which our trajectory length of $N=10^6$
is apparently insufficient.
For the infinite models such as R$\infty$ the motion of the particle is
never completely predictable, but the proportion of collisions that the
particle encounters new scatterers decreases as $1/\log t$ according to
the discussion of Sec.~\ref{s:chaos}.  The measurement of such a slow
decrease is far beyond the capabilities of any feasible implementation of the
Grassberger-Procaccia method.

Since the Grassberger-Procaccia approach does not seem to distinguish
our chaotic and nonchaotic diffusive models, we turn to an alternative
time series analysis method, with a view towards suggesting possible
methods for analyzing future experiments that are able to sample the
dynamics on microscopic time scales.  Such a method cannot possibly
enumerate all possible trajectory segments, as discussed above, but there
are certain types of trajectory segments, namely those that are almost
periodic, that stand out as predictable in a nonchaotic system.  The method
of the next section takes advantage of this property, and in effect
distinguishes chaotic and nonchaotic systems by singling out this subset
of all possible recurrences.

\section{Almost periodic recurrences}\label{s:ISR}
\subsection{Motivation}\label{s:motiv}
We have seen in Sec.~\ref{s:GP} that the variant of the Grassberger-Procaccia
method used by Gaspard et al cannot distinguish between nonchaotic and
chaotic models, for the reason that the time series (either experimental or
numerical) is not long enough to provide a reasonable approximation to the
infinite time limit implied by the method.  

We can attempt to circumvent this difficulty by looking for specific types
of recurrences, as opposed to taking an average over all types of recurring
trajectory segments.  In particular we will show that a promising
candidate for a useful specific recurrence is given by one that corresponds
to an orbit that is almost periodic (again, within a spatial tolerance
$\epsilon$).  We can identify these almost periodic orbits as those that
repeat (within a distance $\epsilon$) their previous motion at equally spaced
intervals of time, hence the name ``Almost periodic recurrences'' (APR).

If we compare the APR approach with the Grassberger-Procaccia (GP) method,
there are a number of similarities and differences.  Both methods
are designed to deduce dynamical properties from time series data.
GP aims to make quantitative statements about the KS entropy, whereas APR
as yet only provides a qualitative statement about chaoticity as expressed
in periodic orbit properties, which are not equivalent to the usual definition
of chaos as positive KS entropy.  Both methods
make use of the probability of recurrences.  GP then
averages over many probabilities, while APR singles out a few 
especially significant recurrences.  This means,
for example that in an intermittent system, where the dynamics switches
irregularly from chaotic to regular behavior, GP measures only the average,
chaotic, dynamics, while APR is sensitive to the lack of chaoticity in
the regular regions of phase space.  In other words, an intermittent system
can have a positive Lyapunov exponent, and also periodic orbits with
nonchaotic properties.  The APR method characterizes such a system as
nonchaotic, which is an appropriate designation with regard to diffusive
properties as we observe in Sec.~\ref{s:esc} below, but which is
inappropriate from the usual point of view, which is that chaos is defined
as positive KS entropy.  An example of such an intermittent system
is given in the final discussion.
 
We now discuss the properties of periodic orbits in chaotic and nonchaotic
systems (particularly our models), before describing the APR method in more
detail and presenting the results.

\subsection{Periodic orbits in chaotic systems}\label{s:CPO}
Periodic orbits are of great importance in chaotic systems.  Although
typical trajectories (as defined by the Liouville or natural measures)
are not periodic, there are many systems for which they are approximated
by periodic orbits when the periodic orbits are dense in the phase
space or on an attractor.  This allows many properties to be computed
from expansions involving periodic orbits~\cite{Cv}.

The properties of the periodic orbits are not directly related to the
other dynamical properties discussed in Sec.~\ref{s:chaos}, yet they do tend
to differ between chaotic and nonchaotic systems, and so they constitute
additional dynamical properties related to chaos.
The properties of the periodic orbits
in the Lorentz gas are of most interest to us, and in some sense
typical for chaotic systems, so we focus on these.  In this and the following
section we examine two relevant properties of periodic orbits, and
sketch proofs of these properties.

\paragraph*{Existence}
There is an easy constructive proof of the existence of periodic orbits in
the Lorentz gas: Choose two disks arbitrarily; the line joining their
centers gives a periodic orbit unless there is a disk interposing; consider
one of the original disks and the interposing disk, and repeat until no disk
interposes.

\paragraph*{Stability}
The periodic orbits are exponentially unstable, that is, almost all
trajectories beginning a small distance $\epsilon$ from the periodic
orbit at time $t=0$ are at a distance $\epsilon e^{\lambda_p t}$ at time
$t$, where $\lambda_p$ is the maximum (here the only positive) Lyapunov
exponent, which depends on the periodic orbit $p$.  This exponential
instability makes it very unlikely for a typical trajectory to remain
near a given periodic orbit for long times.

\subsection{Periodic orbits in nonchaotic systems}\label{s:NPO}
We study here the properties of periodic orbits of our wind-tree models,
which are typical of nonchaotic systems, without attempting to classify
all possible nonchaotic periodic behavior.  For a complementary study of
the periodic orbits in polygonal billiards, see Ref.~\cite{B}.

Not all wind-tree models have periodic orbits.  The argument given above for
the Lorentz gas fails because an orbit connecting two square scatterers is
periodic only if the scatterers have the same orientation (of zero probability
in the randomly oriented model) and the particle moves perpendicular to the
surface of the scatterers (excluded by definition in the fixed orientation
model).  In general, the existence of periodic orbits depends on the
locations and orientations of the scatterers.  We will find that using
generic locations of
scatterers periodic orbits exist in the R$\infty$ model, but not in the
F$\infty$ model.  The periodic models (such as FP4 and RP4) are harder to
treat mathematically, and are not discussed here.

\paragraph*{F model: nonexistence}
The absence of periodic orbits for generic configurations of the F$\infty$
model follows directly from arguments in a proof of Aarnes~\cite{A}.
In order for the particle to return to its starting point, an expression
linear in the positions of the scatterers must vanish.  The coefficients
of the $(x,y)$ position of a scatterer in this expression are integers
depending on which faces the particle hits, and are zero only if the
particle hits opposite faces an equal number of times.  A simple
{\em reductio ad absurdum} argument shows then that there is a scatterer
(perhaps
more than one), the furthest to the top and right (largest value of $x+y$),
which has collisions on only one of its faces (the lower left), and hence
the coefficients corresponding to the position of this scatterer are nonzero.
If the scatterers are randomly placed, no linear combination of the positions
can vanish, so no periodic orbit can exist.  Of course, there are special
configurations of the scatterers (such as at the corners of a square aligned
with the coordinate axes) that allow periodic orbits.
Since there are therefore generically no periodic orbits in the
F$\infty$ model, the remainder of this section refers to the R$\infty$ model.

\begin{figure}
\begin{picture}(280,280)(-50,0)
\put(0,0){\epsfig{file=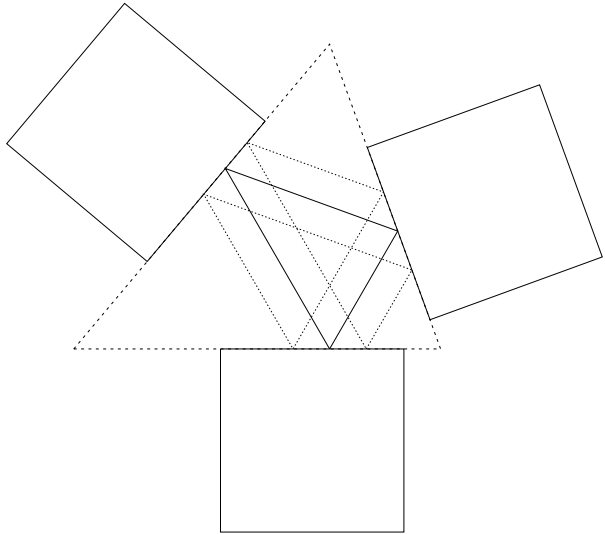, width=280pt}}
\end{picture}
\caption{Construction of periodic orbits in the infinite randomly oriented
wind-tree model R$\infty$.  Any acute angled triangle (dashed), generated
by three scatterers (solid squares), contains a period 3 orbit of minimal
length (solid triangle).  It is surrounded by a family of period 6 orbits
(dotted) that differ from the period 3 orbit by the location of the collisions
but not the direction of the velocities.
\label{f:tri}}
\end{figure}

\paragraph*{R model: existence}
The existence of periodic orbits for generic configurations of the 
R$\infty$ model follows from the observation that any set of configurations
of a finite number of ({\em eg. } 3) scatterers with nonzero (Lesbesgue)
measure includes configurations appearing somewhere in a generic infinite
configuration.  A period 3 orbit can always be found in an acute
triangular billiard by minimizing the total path length, see Fig.~\ref{f:tri}.
If three scatterers outline an acute triangle and are positioned so that
their faces contain the path of minimal length (clearly of nonzero
measure), a period 3 orbit exists.

\paragraph*{R model: stability}
Finally we investigate the stability of periodic orbits in the R$\infty$
model.  The combination of the linear dynamics of the free particle with
length preserving reflections leads to a linear separation of nearby
trajectories.  That is, an initial separation of $\epsilon$ in the
direction of the velocity leads to a separation $\epsilon t$ in the position
after time $t$.  Another way of saying this is that the number of uniformly
distributed trajectories remaining within a distance $\epsilon$ of the
periodic orbit is $\epsilon/t$ for large $t$.  Thus a particle in the R$\infty$
model is quite likely to spend a long time near a periodic orbit, in contrast
to in the Lorentz gas.
 
\subsection{Details and results}
Now we return to our method of Almost Periodic Recurrences for
distinguishing chaotic properties of diffusive systems from a time series.
This consists of counting the number of almost periodic sequences in the
time series, thus hoping to exploit the different properties of periodic
orbits in chaotic and nonchaotic systems as discussed in the previous two
sections.

As in the Grassberger-Procaccia method we begin with a time series containing
$10^6$ positions of the particle $(x_i,y_i)$ spaced at a time interval
$\Delta t=1$, see Sec.~\ref{s:time}.
Analogous to Eq.~(\ref{e:dist}) we define a distance between
two segments of length $T$ (the period of an almost periodic orbit) that
begin at points $i$ and $j$ on the trajectory,
\begin{equation}\label{e:dT}
d_T(i,j)=\max_{k=0}^{T-1}\sqrt{(x_{i+k}-x_{j+k})^2+(y_{i+k}-y_{j+k})^2}
\end{equation}

\begin{figure}
\begin{picture}(330,470)(-50,0)
\put(0,320){\epsfig{file=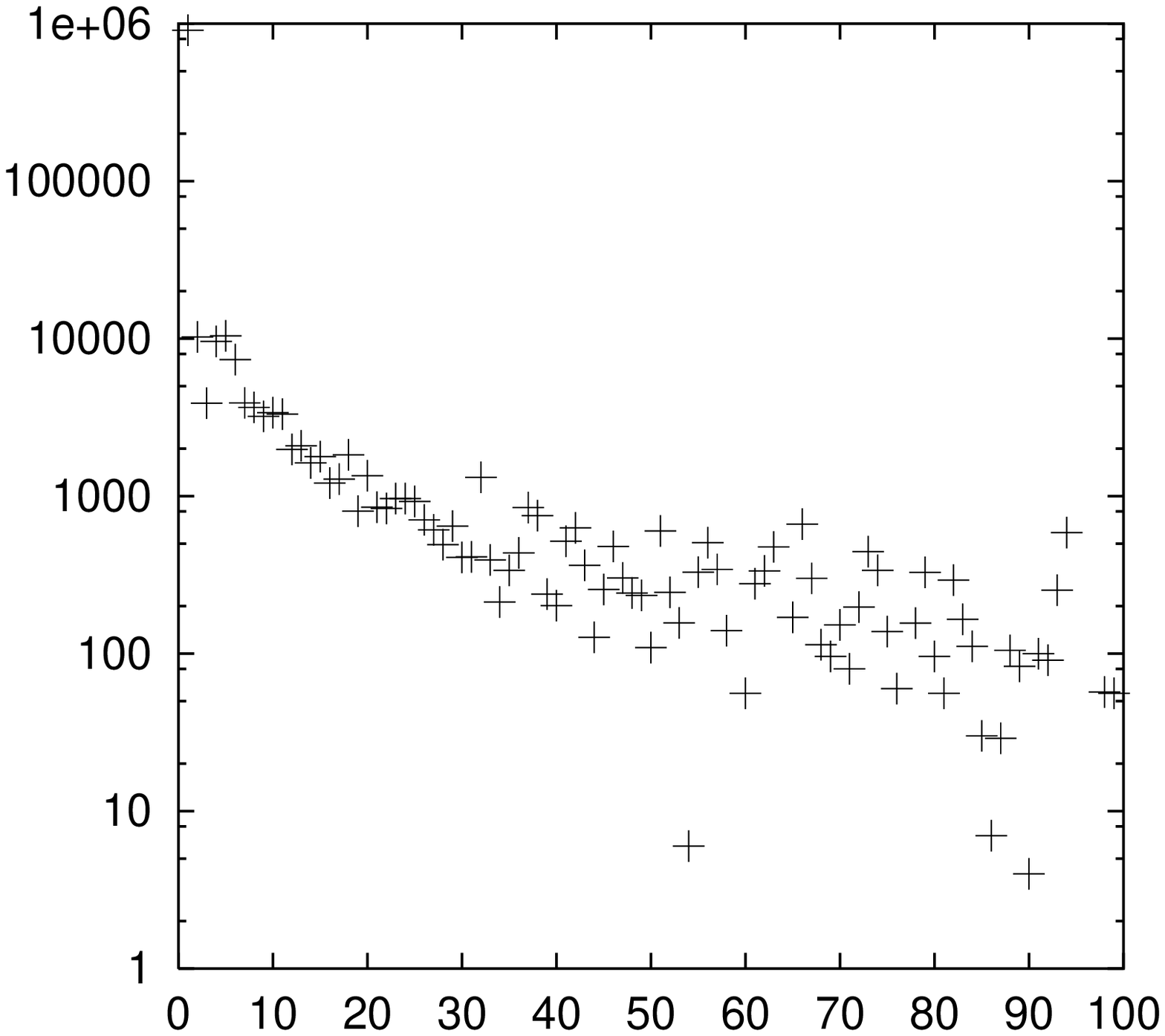, width=160pt}}
\put(150,320){\epsfig{file=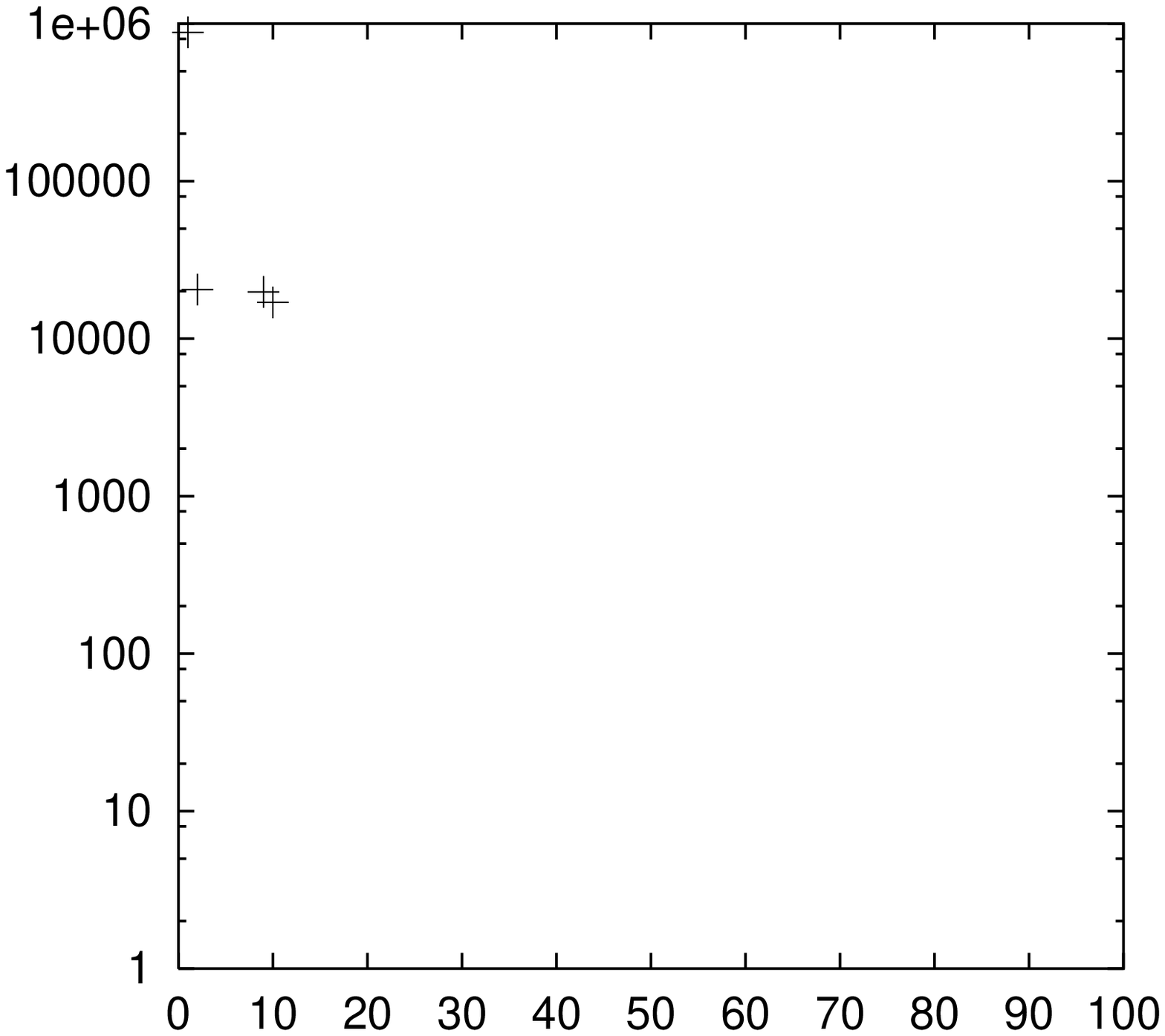, width=160pt}}
\put(0,170){\epsfig{file=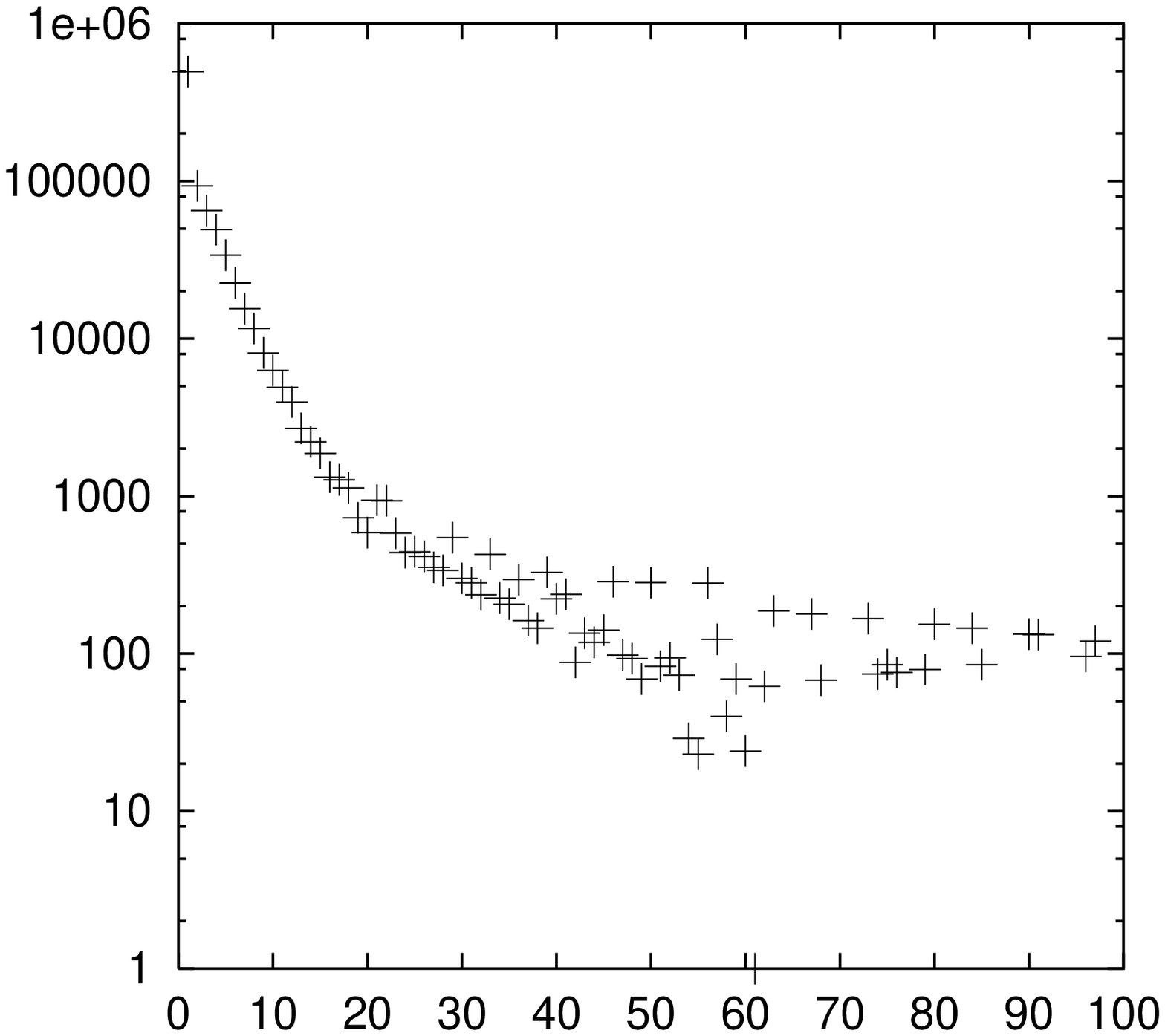, width=160pt}}
\put(150,170){\epsfig{file=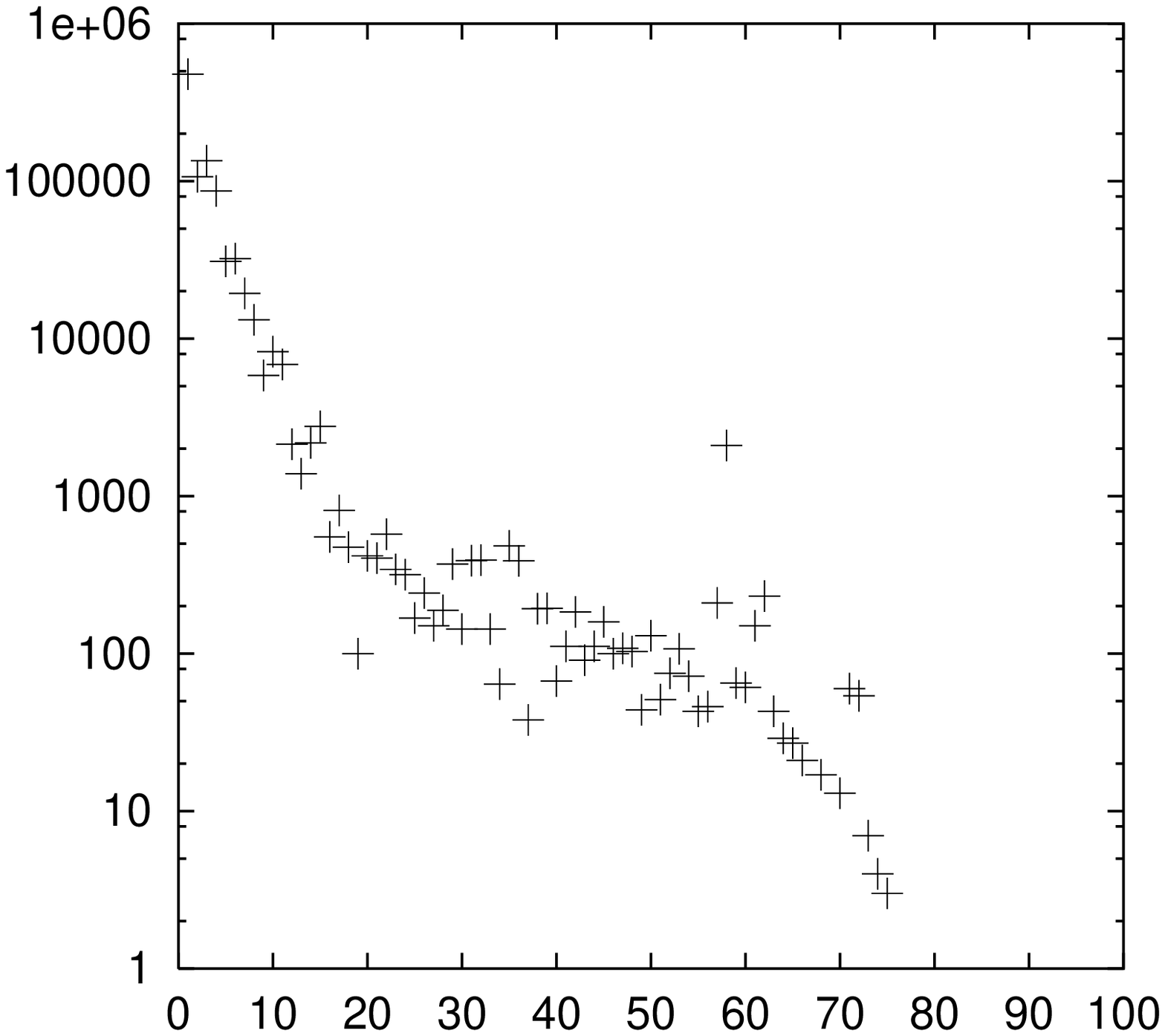, width=160pt}}
\put(0,20){\epsfig{file=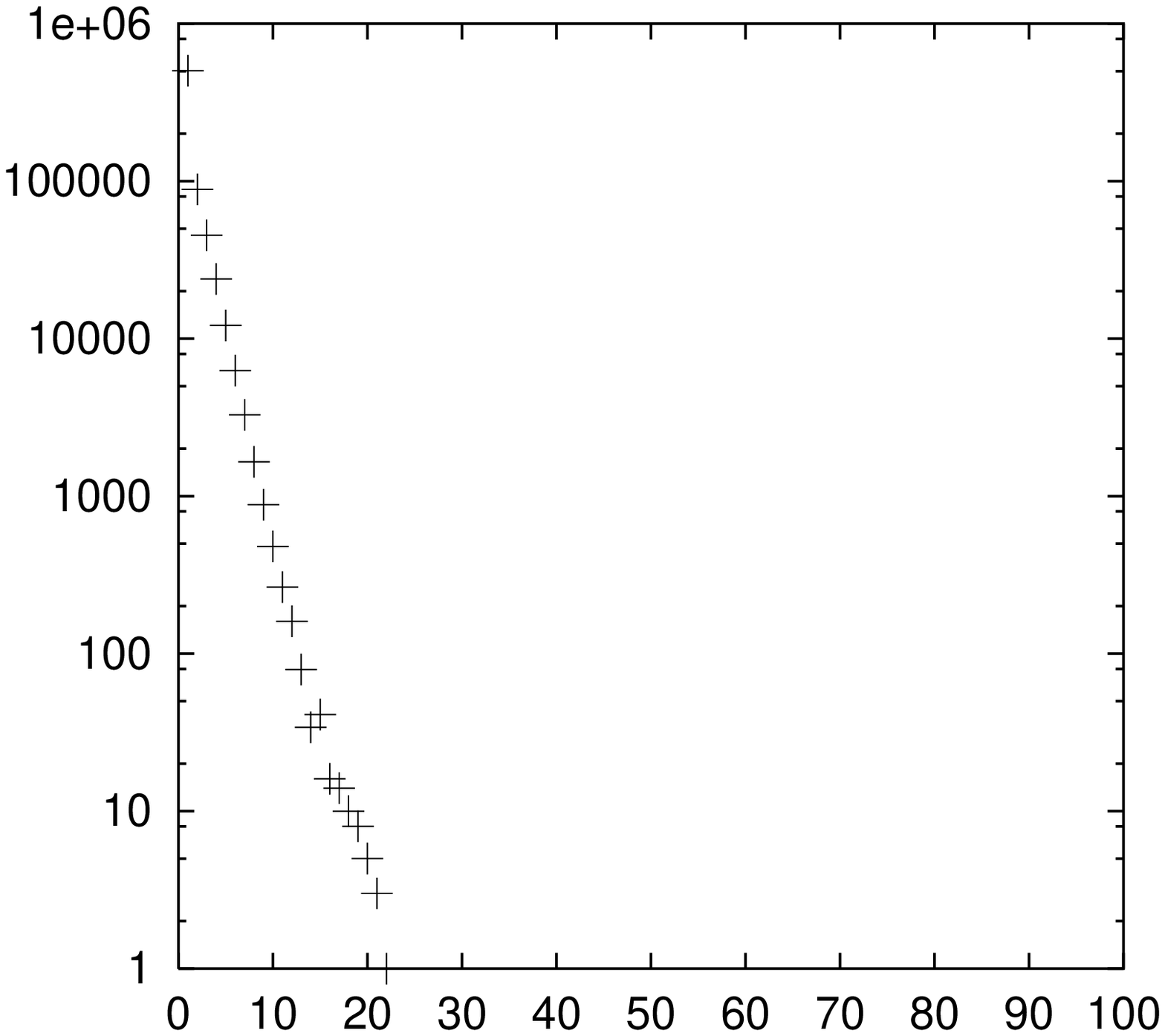, width=160pt}}
\put(150,20){\epsfig{file=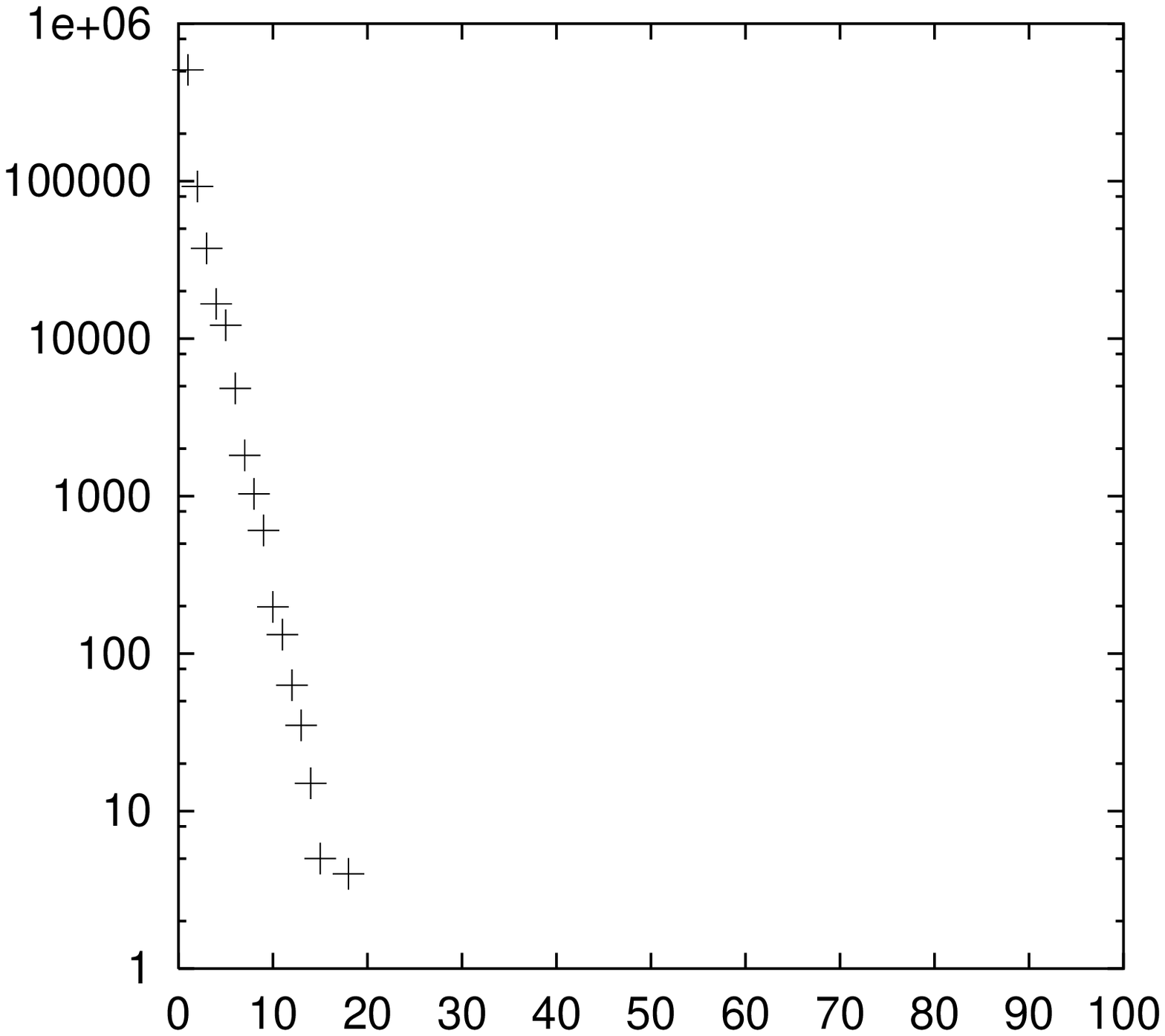, width=160pt}}
\put(85,10){$T$}
\put(-15,95){$N_T$}
{\large
\put(-10,430){F}
\put(-10,280){R}
\put(-10,130){L}
\put(85,460){$\infty$}
\put(235,460){P4}}
\end{picture}
\caption{Results of the APR method, showing the number of times in which
a periodic orbit of a given length $T$ was repeated within a tolerance
$\epsilon=1$.  The models (see Sec.~\protect\ref{s:not}) are F (upper),
R (middle) and L (lower), each with $\infty$ (left) and P4 (right).  
The chaotic Lorentz models show a clear exponential decrease
with $T$, while the nonchaotic models exhibit distinctly different behavior
depending on the properties of their (almost) periodic orbits; see the text. 
\label{f:ISR}}
\end{figure}

Using Eq.~(\ref{e:dT}) we compute the number of initial points $i$ for which
the orbit repeats within a tolerance $\epsilon$,
\begin{equation}
N_T(i)=\#_i[d_T(i,i+T)<\epsilon]
\end{equation}
where we use $\epsilon=1$ for the results presented in Fig.~\ref{f:ISR}.

It is clear that the chaotic Lorentz models, with an exponential decay of
$N_T$ with $T$ can be easily distinguished from the nonchaotic F and R models
which mostly have a slower and noisier decay, so we have achieved our objective
of finding a time series analysis method that can distinguish between time
series of chaotic and nonchaotic diffusive systems.  It remains for us
to explain the difference, and the plots obtained using the properties of
the periodic orbits in these systems.  For that, we note that
there are two types of (almost) periodic orbits that might appear in the
expression for $N_T$: An orbit of length $T$ that has been (almost) repeated,
and a shorter orbit of length $T/n$ that has appeared $2n$ times, so that
the total time is equal to $2T$. 

\paragraph*{The L models} 
In the Lorentz gas, the probability of remaining near a periodic orbit
for a time $T$ is of order $e^{-\lambda_p T}$ (see Sec.~\ref{s:CPO}), where
$\lambda_p$ is the maximum (here the only) positive Lyapunov exponent, and
depends on the periodic orbit $p$.  This is why both the infinite and
periodic Lorentz gases give an exponential form in Fig.~\ref{f:ISR}.
Because $N_T$ can be due to long orbits or repeats of short orbits, the
exponential form depends on both (1) the exponential escape from short orbits
with many repetitions and (2) the exponential instability of orbits with
length, that is, that $\lambda_p$ does not approach zero for long orbits.
\footnote{
There is a minor technical difficulty associated with the random Lorentz
gas: It is possible due to the random placement of the scatterers to have
arbitrarily long free paths between collisions, leading to the possibility
that $\lambda_p$ could become arbitrarily small.  However, the probability
of a long free path is also exponentially small, so the result is still
exponential. }

\paragraph*{The R models}
In the randomly oriented wind-tree models, the probability of remaining
near a periodic orbit for a time $T$ is of order $1/T$ (see Sec.~\ref{s:NPO}).
This form is not immediately apparent in
Fig.~\ref{f:ISR} because orbits (particularly
those surviving for long times) can be counted more than once due to
contributions from an interval of different initial points $i$.  Both the
R$\infty$ model and the RP4 model continue beyond $T=100$ due to increasingly
rare long orbits and many repeats of shorter orbits.  For example, the
almost periodic orbit giving rise to the conspicuous point with period
$T=58$ in the RP4 model is repeated many times, also contributing to multiples
of the period, double $T=$114,115; triple $T=$172,173; and so on up to
$T=$1033,1034.  It is clear that the
dynamics is completely different to that of a chaotic system.

\paragraph*{The F models}
In the fixed oriented wind-tree models, there are no periodic orbits
(see Sec.~\ref{s:NPO}).  This means that any contribution to $N_T$ is due
to orbits that are close to periodic.  The infinite model F$\infty$ contains
many arrangements of a few scatterers that come arbitrarily close to generating
a true periodic orbit, so the plot is similar to that of the R models.
The periodic model FP4, in contrast, has only a few short orbits that are
at all close to periodic, and none that can be repeated to give further
contributions at higher values of $T$, or these would be observed in the
figure.

What we have presented here barely scratches the surface
of the Almost Periodic Recurrences method and its variants.  It would be
quite easy to search for 3 or more occurrences of an (almost) periodic orbit,
or to distinguish between a single repeat of a long orbit and many repeats
of a short orbit.  We have restricted the presentation here to a demonstration
of the method as a means to distinguish time series of chaotic and nonchaotic
diffusive systems, but we hope that it will be useful in more general contexts.

\section{Absorbing boundary conditions}\label{s:esc}
The case of absorbing boundary conditions provides a rich context for
illuminating the differences between chaotic and nonchaotic diffusion.
This is because the diffusion equation with absorbing boundary conditions
leads naturally to an exponential escape process, while the periodic orbits
of the randomly oriented wind-tree model require a power law escape
process (see Sec.~\ref{s:NPO}), thus violating the diffusion equation.

We consider here the probability that a single randomly placed particle will
remain in an open system\footnote{We describe systems with absorbing
boundaries as ``open''.}
for a given time, or equivalently, the number of identical
noninteracting particles remaining in the system
after the same time.  For chaotic dynamics
the escape rate formalism of Gaspard and Nicolis~\cite{Gaspard,GN,GD} uses
the escape rate (defined below) to connect the diffusion coefficient to
dynamical properties such as the KS entropy and the
positive Lyapunov exponents.
For nonchaotic systems the escape process is qualitatively different
depending on the properties of the periodic orbits as we have noted above,
and the escape rate formalism must be generalized, if it is to make any sense
at all.  We now describe the escape rate formalism as it applies to chaotic
systems, discuss the results for our models, and then attempt to generalize
the formalism to include nonchaotic systems.

A macroscopic description of escape on a square (for simplicity; other
geometries are analogous) is given by the diffusion equation
(\ref{e:diff}) for the particle density $P({\bf x},t)$, together with
the boundary condition $P=0$ along the lines $x=0$, $y=0$, $x=L$ and
$y=L$.  The general solution is
\begin{equation}\label{e:diffsol}
P({\bf x},t)=\sum_{m,n=1}^{\infty}a_{m,n}\sin(m\pi x/L)\sin(n\pi y/L)
e^{-\gamma_{m,n}t}
\end{equation}
with the decay rate
\begin{equation}
\gamma_{m,n}=\frac{D\pi^2}{L^2}(m^2+n^2)\label{e:gmn}
\end{equation}
At long times, the solution is dominated by the slowest decaying mode,
corresponding to the escape rate
\begin{equation}\label{e:er}
\gamma=\gamma_{1,1}=\frac{2D\pi^2}{L^2}
\end{equation}
Note that the escape of particles is exponential in~(\ref{e:diffsol}).
The escape rate formalism equates this macroscopic escape rate
$\gamma$ with the (exponential) escape rate of a (microscopically)
chaotic system, which
is related to its KS entropy $K$ and the sum of its positive
Lyapunov exponents $\sum\lambda_+$ (here at most a single positive
Lyapunov exponent) by~\cite{ER}
\begin{equation}\label{e:erf}
\gamma=\sum\lambda_+-K
\end{equation}
the ``escape rate formula'' which generalizes Pesin's formula
(Sec.~\ref{s:chaos}) to open systems.  Not only can the escape rate of
chaotic systems be calculated from periodic orbit theory~\cite{Cv}, but,
as we will show below, there is an intimate connection between periodic
orbits and the escape process in nonchaotic systems. 

\begin{figure}
\begin{picture}(300,320)(-50,0)
\put(0,20){\epsfig{file=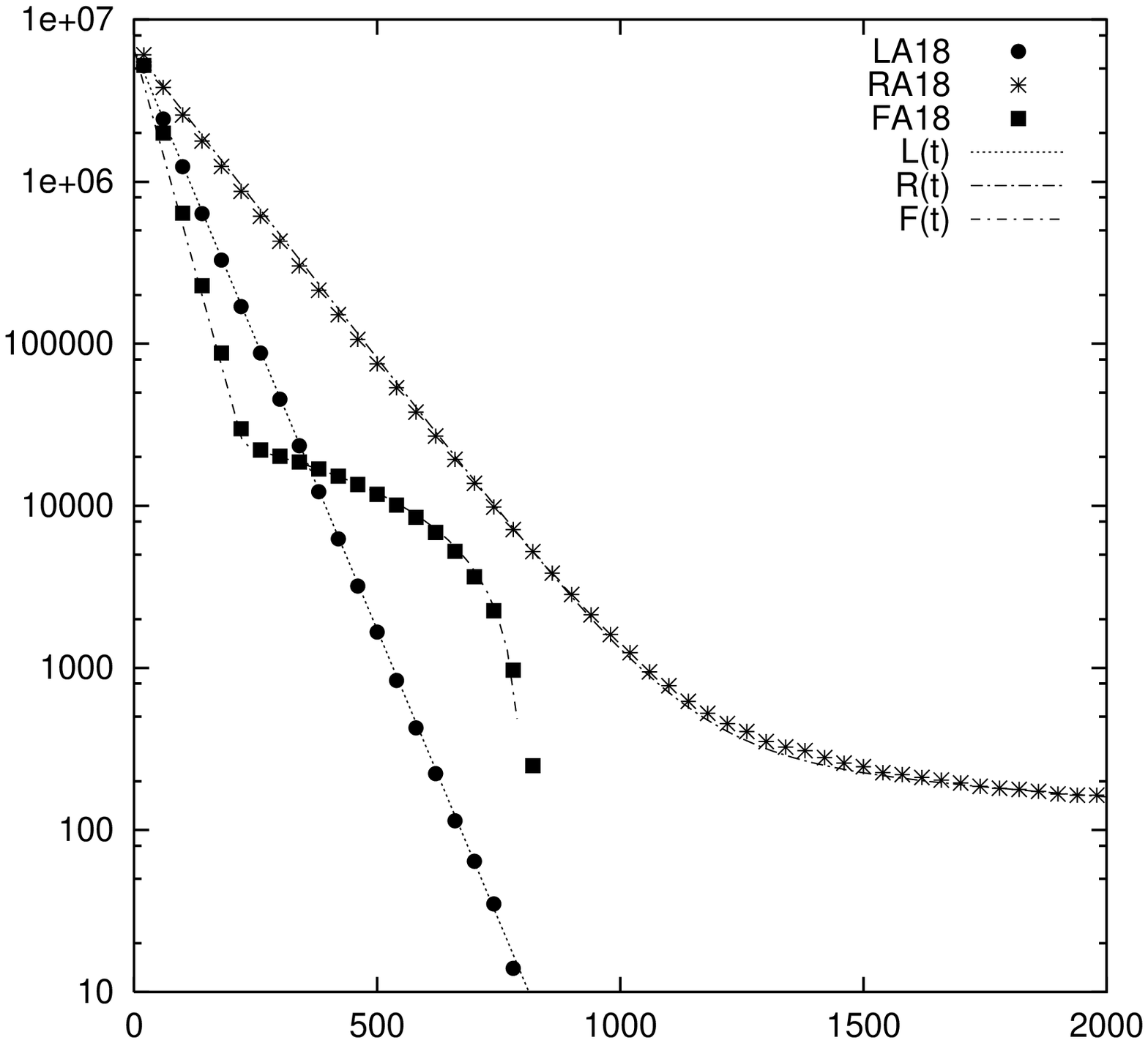, width=300pt}}
\put(165,10){\large $t$}
\put(-15,165){\large $N(t)$}
\end{picture}
\caption{The decay of the number of particles with time when the boundaries
are absorbing for the Lorentz gas LA18 (circles), randomly oriented
wind-tree model RA18 (stars) and the fixed oriented wind-tree model
FA18 (squares).  The decay is exponential for short times for all three
models as predicted by the diffusion equation, and model-dependent
for long times.  The LA18 model remains exponential, while the RA18 model
switches to a $1/t$ form and the FA18 model switches to a linear $t_c-t$ form.
The results are well described by the functions $L(t)$,
$R(t)$ and $F(t)$ respectively, see Eqs.~(\protect\ref{e:L}-\protect\ref{e:F})
\label{f:esc}}
\end{figure}

\begin{figure}
\begin{picture}(300,320)(-50,0)
\put(0,20){\epsfig{file=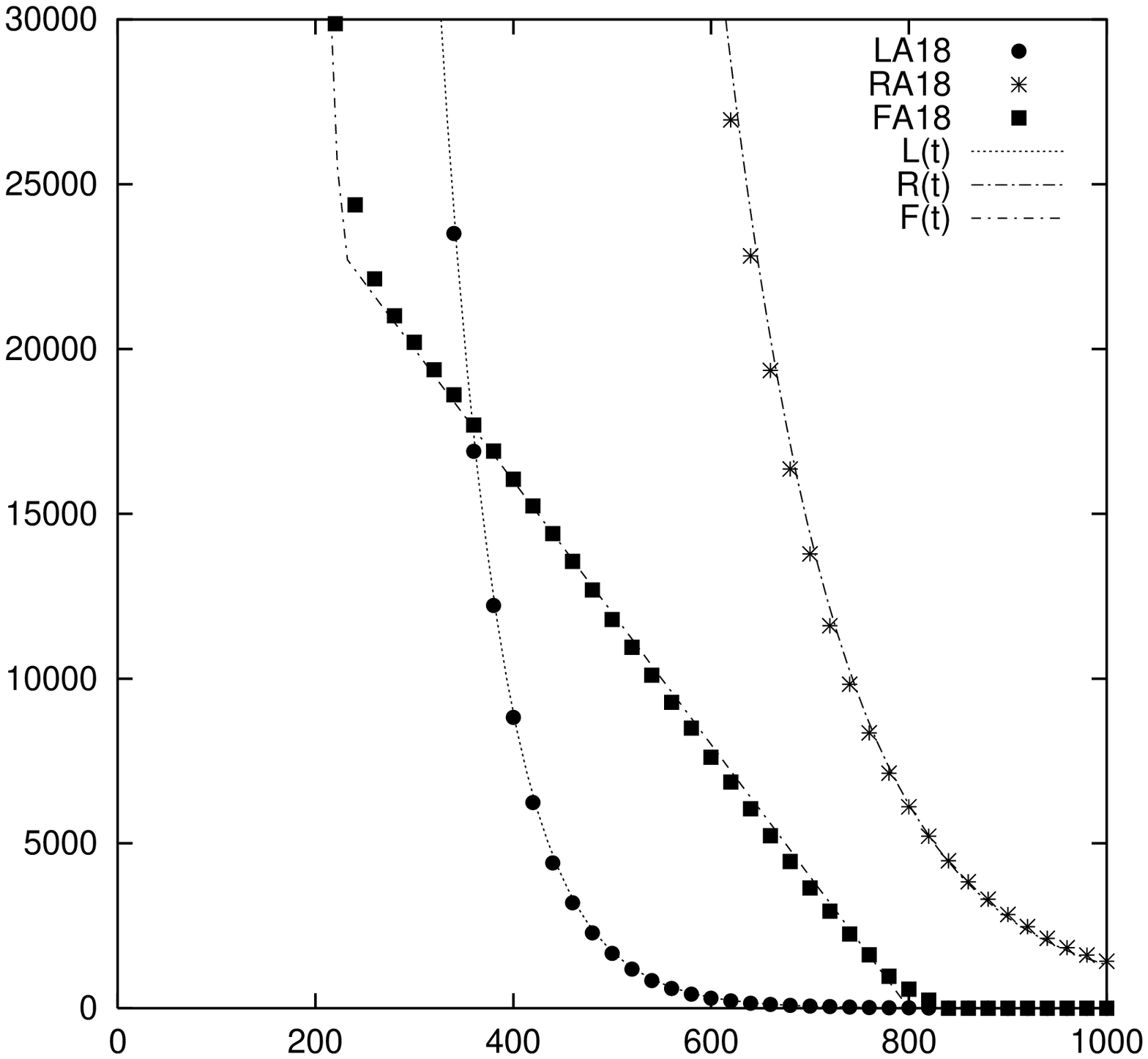, width=300pt}}
\put(165,10){\large $t$}
\put(-15,165){\large $N(t)$}
\end{picture}
\caption{Fig.~\protect\ref{f:esc} plotted with a linear vertical axis
to show the linear form of the escape in the FA18 model.
\label{f:esc2}}
\end{figure}

We now present our numerical results, from which we learn how the above
theory for chaotic systems is modified in the nonchaotic case.  As described
in Sec.~\ref{s:comp}, the arrangement of scatterers used is the same as in
the periodic case with $L=20$; the absorbing boundaries are one distance
unit from the edge of the periodic cell, leading to an open system of
size $L=18$.  As before we use fixed oriented squares (F), randomly oriented
squares (R) or circles (L).  We denote absorbing boundary conditions by
A, so the full notation for these models is [F,R,L]A18.
We place $10^7$ particles uniformly (without overlapping the scatterers)
in the square of size $L=18$ and compute the number of particles remaining
in the system as a function of time, see Figs.~\ref{f:esc},~\ref{f:esc2}.

The chaotic Lorentz gas exhibits exponential decay as predicted by the
diffusion equation for all times, but the nonchaotic wind-tree models
deviate from exponential decay at late times.  The results are
well described by the empirical expressions
\begin{eqnarray}
L(t)&=&6.6\times10^6\times e^{-0.0165t}\label{e:L}\\
R(t)&=&6.6\times10^6\times e^{-0.0088t}+3.2\times10^5/t\label{e:R}\\
F(t)&=&\left\{\begin{array}{ll}
6.6\times10^6\times e^{-0.025t}&t<226\\
40(800-t)&226<t<800\\
0&t>800
\end{array}\right.\label{e:F}
\end{eqnarray}
for the L, R and F models respectively over the range of times considered.
The initial exponential decay may be compared with
Eqs.~(\ref{e:diffsol}-\ref{e:er}) to give an effective diffusion coefficient.
We find $D_L=0.27$, $D_R=0.14$ and
$D_F=0.41$, which are certainly consistent with the results of the mean
square displacement, Tab.~\ref{t:diff} in Sec.~\ref{s:diff}, given that
Eqs.~(\ref{e:diffsol}-\ref{e:er})
depend on the macroscopic diffusion equations, and so require a large
system $L\rightarrow\infty$ limit.  The coefficient $6.6\times 10^6$ is
obtained by matching Eq.~(\ref{e:diffsol}) to the uniform probability
density of the initial conditions; this comes to $64 N(0)/\pi^4$ with
$N(0)=10^7$ as the initial number of particles in the system.  From this
close agreement, we conclude that the early escape is well described by the
diffusion equation with the same diffusion coefficient that appears in the
mean square displacement of Sec.~\ref{s:diff}; there is no trace of
non-diffusive behavior at short times.

\paragraph*{The R model}
We now discuss the long time behavior of the nonchaotic models.
The randomly oriented wind-tree model has a $1/t$ decay in~(\ref{e:R})
due to its
periodic orbits; particles that are initially close to a periodic orbit
lead to this power law as described in Sec.~\ref{s:NPO}.  The coefficient
gives an estimate of the density of periodic orbits; $3.2\times 10^5$ is
quite small compared to the number of particles, $10^7$, so periodic orbits
are relatively rare in a sense that is difficult to define precisely.
In terms of the escape rate formalism, a power law decay corresponds to
an escape rate $\gamma=0$ which trivially satisfies~(\ref{e:erf}), but
yields no information about the diffusion coefficient $D$, that is,
(\ref{e:er}) is not satisfied.

\begin{figure}
\begin{picture}(350,350)(-50,0)
\put(0,0){\epsfig{file=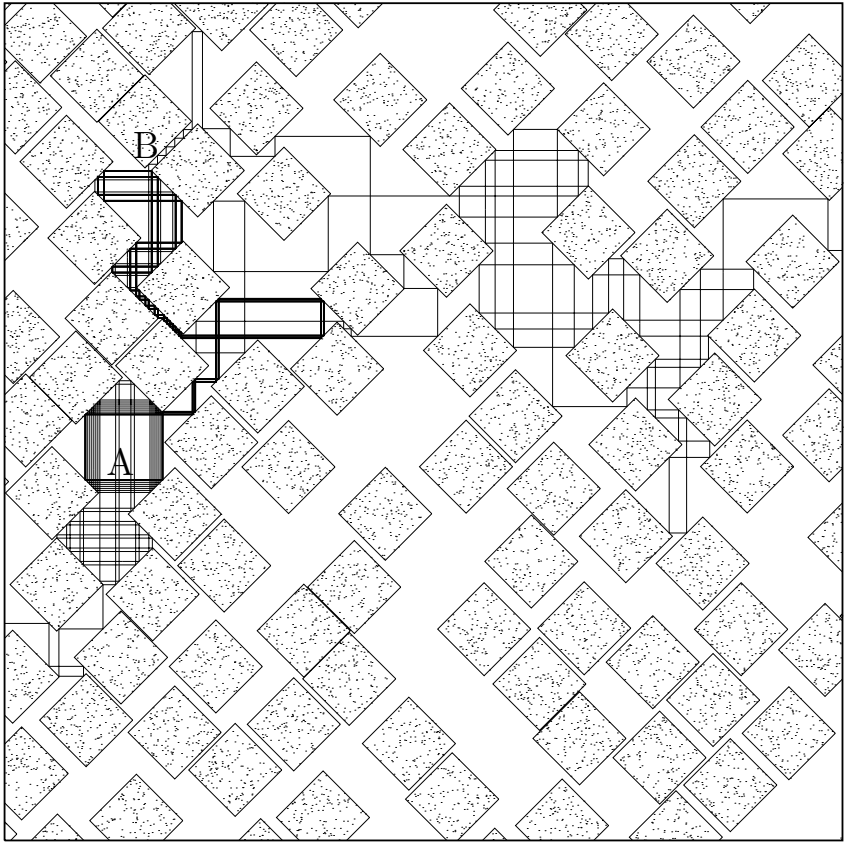, width=350pt}}
\put(55,150){\Large A}
\put(65,270){\Large B}
\end{picture}
\caption{A long lived orbit (total time $t\approx800$) in the FA18 model.
Its length is mainly
due to two almost periodic orbits, a square (A), and a longer
orbit that is so close to periodic that it appears as a pair of
thick lines on the scale of this plot, extending from A to B.
The linear form of Fig.~\protect\ref{f:esc2} indicates that a few such
orbits are responsible for most of the particles remaining in the system
beyond $t=226$, see the discussion in the text.
\label{f:traj}}
\end{figure}

\paragraph*{The F model}
The fixed oriented wind-tree model shows a complete escape of all the
particles in a finite time, as might be expected from the lack of periodic
orbits. The dramatic transition from
the initial exponential decay to a much slower (at first) linear decay
is somewhat
surprising.  We interpret this as follows:  While there are no exactly
periodic orbits in the F model, it is possible for the particle to remain
in an orbit that is nearly periodic for some time, allowing the particle
to remain in the system longer than the exponential decay would predict.
If there is a bundle of trajectories that survive for just 800 time units,
then this would lead to a linear law because particles are initially evenly
distributed along the bundle of trajectories.  If there was, in addition,
a large bundle of trajectories that survive for, say 500 time units, there
would be a kink in the plot, with a sharp decrease in gradient at $t=500$.
The observation that the curve is close to linear (see Fig.~\ref{f:esc2})
thus implies that the
lengths of these long living trajectories are strongly peaked around
800 time units.  One such long lived orbit is depicted in Fig.~\ref{f:traj};
its length is seen to be due to two almost periodic orbits.
The complete escape of the particles corresponds to
an escape rate $\gamma=\infty$, which flagrantly violates both~(\ref{e:er})
and~(\ref{e:erf})
and yields no information about the diffusion coefficient $D$.

We conclude from these observations that, although the wind-tree models
satisfy the diffusion equation when there are no boundaries, that is, the
two time distribution function is of the correct Gaussian form, such a
macroscopic
approximation is not valid when there are absorbing boundary conditions.
This means we cannot determine the diffusion coefficient in the same way
as for the Lorentz gas, Eqs.~(\ref{e:diffsol}-\ref{e:er}),
that is, from
\begin{equation}
D=\frac{1}{2\pi^2}\lim_{L\rightarrow\infty}L^2\lim_{t\rightarrow\infty}
\frac{1}{t}\ln\frac{P_L(t)}{P_L(0)}
\end{equation}
because the infinite time limit leads to zero or infinity as explained above.
Here $P_L(t)$ is the probability of a particles remaining in a system of
size $L$ for time $t$; this is obtained by observing many independent
particles, and taking the limit of an infinite number of particles.

We can, however, attempt to calculate the diffusion coefficient in the
finite wind-tree models by taking the $t\rightarrow\infty$ limit at the
same time as the $L\rightarrow\infty$ limit.  This is because
the diffusion equation is a good approximation even in the
open case as long as the time is not too long, and the decay remains
exponential.  For example, assuming that the
``density of periodic orbits'' in the RA model is independent of $L$ (which
seems to be the case numerically), the transition time from exponential to
power law decay, as determined by comparing the two terms in
Eq.~(\ref{e:R}) using Eq.~(\ref{e:er}) is of order $t_{tr}\approx L^2\log L$.
This should be compared
with the time scale for which the slowest decaying mode of the diffusion
equation dominates, $t_D\approx L^2$, see~(\ref{e:gmn}).  Thus, for a given
$L$, there is a narrow range in time $t_D\ll t\ll t_{tr}$ in which the decay is
exponential, and in which a limit can be taken to obtain $D$ from the escape
process.  For example we can combine the $L$ and $t$ limits by setting $L=u$
and $t=u^2\sqrt{\log u}$, 
\begin{equation}\label{e:narrow}
D=\frac{1}{2\pi^2}\lim_{u\rightarrow\infty}\frac{1}{\sqrt{\log u}}
\ln\frac{P_u(u^2\sqrt{\log u})}{P_u(0)}
\end{equation}
We expect this relation to hold for all the models [L,R,F]A.  We emphasize
that Eq.~(\ref{e:narrow}), if it is mathematically valid, still does not
provide a practical method for computing the diffusion coefficient beyond
the estimates we gave following Eq.~(\ref{e:F}).  This is because the
range of time scales $t_D\ll t\ll t_{tr}$ grows so slowly with $L$ that
extremely large system sizes are required for more precise estimates of $D$.

We can also turn the argument around, and suggest that the escape at
long times of finite systems may be a good experimental technique for
detecting (or ruling out at some level) nonchaotic microscopic
dynamics.  In any case, the range of validity of the diffusion equation
as a macroscopic description of nonchaotic systems is restricted.

\section{Discussion and open questions}\label{s:conc}
We have explored the connections between microscopic chaos and diffusion.
On a superficial level, the details of the microscopic dynamics seem to
have little effect on the diffusion process: Gaussian diffusion is
observed in our RP4 model, which has scatterers with flat sides (hence
no exponential separation of nearby trajectories) and is periodic (hence
there is no source of randomness from a disordered environment).  In addition,
the Grassberger-Procaccia method of time series analysis cannot distinguish
our chaotic and nonchaotic diffusive models because it would require
impractically long data sets.  On a deeper level, however, the subtle
differences between chaotic and nonchaotic diffusion are quite apparent
if you know where to look: Our time series analysis method based on
periodic orbits has no trouble distinguishing between the chaotic and
nonchaotic models, and the long time behavior of escape from an open
system is also determined by the properties of the chaotic versus nonchaotic
periodic orbits.

In the light of our results, we can make a few concrete suggestions for
an experimental determination of the chaotic or nonchaotic properties in
a diffusive system.  Firstly, as we remarked in Ref.~\cite{DCvB}, it is
necessary to make measurements on the distance and time scales relevant to
the microscopic dynamics.  In our models, this question is simplified by the
fact that all microscopic time scales are of order one in our units,
Sec.~\ref{s:time}.  Once this is achieved, one could use an
approach based on the method of Almost Periodic Recurrences, which searches
for periodic behavior; alternatively the late time decay of particles from
an open system can also elicit the chaotic or nonchaotic nature of the
dynamics.

This raises again the question of exactly what dynamical properties would
be measured; there are systems with the positive KS entropy of
the Lorentz gas as well as the power law unstable periodic orbits of the
wind-tree models.  We will now make some remarks about such systems,
although we make no attempt to enumerate all degrees and classes
of chaotic dynamics.

Our example in this discussion is a model containing both circular and
randomly oriented square scatterers.  The circular scatterers lead, as in
the Lorentz gas, to a positive Lyapunov exponent, while the square scatterers
lead, as in the randomly oriented wind-tree model, to power law unstable
periodic orbits, using the same argument as in Sec.~\ref{s:NPO}.  We have
not simulated such a model numerically; however the arguments used in the
previous sections may be applied, leading to the following predictions:

\begin{description}
\item{Sec.~\protect\ref{s:conn}:} There will be a positive diffusion
coefficient with a Gaussian distribution function, as in both the L and
R models.
\item{Sec.~\protect\ref{s:GP}:} The Grassberger-Procaccia method will
describe the dynamics as chaotic, due to a positive Lyapunov exponent
and hence positive KS entropy, as in the L models.
\item{Sec.~\protect\ref{s:ISR}:} The APR method will describe the dynamics
as nonchaotic, due to the power law unstable periodic orbits, as in the
R models.
\item{Sec.~\protect\ref{s:esc}:} Such a system with absorbing boundaries
will exhibit power law escape at long times, again due to the periodic
orbits, as in the R model.
\end{description}

The different conclusions reached by the GP and APR methods exemplify the
fact that these methods are based on different dynamical properties; for
diffusion in an open system, it is clear that the mixed model is most
similar to the R model, and hence that the APR method distinguishes the
relevant dynamical property in this case.

We conclude with some open questions and possibilities for further work.
In terms of {\em microscopic chaos}, we have studied only a few
models.  We note that there are examples of lack of
chaoticity we have not considered, such as the coexistence of chaotic
and nonchaotic regions found in KAM theory.  In terms of {\em diffusion},
we note that in the
light of Sec.~\ref{s:esc} the diffusion equation is not a complete
macroscopic description of nonchaotic diffusion.  In terms of {\em time series
analysis}, our APR method, while sufficient for our purposes here, could
be much more developed and applied.  With regard to {\em periodic
orbit theory}, the methods developed for exploiting the importance of
periodic orbits to compute properties of chaotic systems~\cite{Cv} do
not apply to nonchaotic systems, and yet we have seen that periodic orbits
also play an important role in nonchaotic systems.

We note that our infinite models, while sharing some of
the properties of corresponding finite billiards, appear to fall outside the
domain of current mathematics.  For example, a recent study of infinite
billiard systems~\cite{T} is mostly restricted to cases with finite areas
that are
finitely connected; our models obey neither of these conditions.  It would
be interesting to see if it is possible to rigorously determine the status
of our models, particularly their ergodic properties, rate of decay of
correlations and Kolmogorov-Sinai entropy within the framework of some
mathematical theory of infinite nonchaotic systems.  More generally,
a mathematical understanding of statistical mechanics including the
thermodynamic limit naturally leads to the study of infinite systems.

Finally, we remark that all the models we have considered --- with the
exception of one (FP4) --- are diffusive, limiting our investigation
as to the presence of microscopic chaos or not to such systems.  In addition,
the precise role played by microscopic chaos --- as represented by the
Lyapunov exponents --- and ``macroscopic chaos'', as embodied by the
randomly placed scatterers for the existence of a diffusion process and
the value of the diffusion coefficient, remains open.  A similar but
more complicated situation obtains when diffusion of momentum (viscosity)
or energy (heat conduction) and other transport processes are considered.

\section*{Acknowledgments}
First of all we want to acknowledge that H. van Beijeren suggested the
wind-tree model as an important alternative to the Lorentz gas for the
study of microscopic chaoticity and diffusion, which led to the work
of Ref.~\cite{DCvB}
and this paper.  We thank him also, along with L. Bunimovich, J. R. Dorfman,
P. Gaspard and I. Procaccia for stimulating and helpful discussions.
This work was supported by the
Engineering Research Program of the Office of Basic Energy Sciences of
the US Department of Energy under contract \#DE-FG02-88-ER13847.

\appendix
\section{The mean free time}
We give here a calculation of the mean free time for our models quoted in
Sec.~\ref{s:time}.  Our derivation is due to Chernov~\cite{Chernov},
extended to the fixed oriented model, and with a technical caveat
for the infinite models.  Refer to Sec.~\ref{s:comp} for the definitions
of $R$, $L$, $N$ and $\rho$.

The mean free time $\bar\tau$ is equal to the mean free
path, since the velocity is one.  The mean free time is known exactly
for billiard systems\cite{Chernov}, which include the [R,L]P models
discussed here.  We make a minor extension to allow the FP model (which
differs because we do not want to allow all velocity directions).
We expect that the formula for the mean
free time would still hold in the infinite models on
physical grounds, but we cannot justify this mathematically.
Briefly, the argument in Ref.~\cite{Chernov} observes that the total volume
of phase space $V$ can be computed in two different ways.

One expression for the volume of phase space is as an integral over the area
of the billiard, giving
$V=4A$ for the original wind-tree model with fixed orientations, and
$V=2\pi A$ for random orientations or the Lorentz gas.  Here, $V$ is the
volume of phase space, including both position and velocity, while $A$
is the area accessible to the point particle.
For our infinite billiards we think of
a large but finite periodic system of length $L$, and formally take the limit
$L\rightarrow\infty$ at the end of the calculation.  Since the area $A$
is equal to $L^2(1-\rho)$, we have $A=L^2/2$ here as $\rho=1/2$.
The prefactor $4$ comes from
the four possible wind directions, and the $2\pi$ from integrating over all
possible directions.  Putting these expressions together we have
$V=4L^2(1-\rho)$ for fixed oriented squares and
$V=2\pi L^2(1-\rho)$ for randomly oriented squares or circles.

An alternative expression for the volume of phase space is an integral
over the boundary of the billiard, where the contribution from each point
is given by the free path/time $\tau$.
In this way we find $V=\sqrt{2}\bar\tau P$ for fixed
oriented squares and $V=2\bar\tau P$ for randomly oriented squares or circles.
Here $P$ is the length of the boundary, that is, the total perimeter of all
the scatterers, $\bar\tau$ is the {\em mean} free path/time,
and the numerical prefactor is the integral over the component of velocity
perpendicular to the boundary.  For the wind-tree models we have
$P=4\sqrt{2}N$ where $N$ is the number of scatterers, so
$P=2\sqrt{2}\rho L^2$;  for the Lorentz gas, we have
$P=2\pi RN=\sqrt{2\pi}\rho L^2$.

Comparing the expressions for $V$ in both calculations, we find
$\bar\tau=(1-\rho)/\rho=1$ (using our value of $\rho=1/2$) for the case
of fixed oriented squares,
$\bar\tau=\pi(1-\rho)/(2\sqrt{2}\rho)=\pi/(2\sqrt{2})\approx 1.111$ for
the case of random oriented squares, and
$\bar\tau=\sqrt{\pi/2}(1-\rho)/\rho=\sqrt{\pi/2}\approx 1.253$ for the
Lorentz gas.

\end{document}